\documentclass[twocolumn,showpacs,prb,footinbib,a4paper,superscriptaddress,floatfix]{revtex4}

\usepackage{amssymb,amsmath}
\usepackage{graphics}
\usepackage{dcolumn}
\usepackage{bm}
\usepackage{epsfig}

\newcommand{\expect}[1]{\langle #1 \rangle}

\newcommand{\SEs}{{\rm self-energies}\ }
\newcommand{\NE}{{\rm non-equilibrium}\ }
\newcommand{\MB}{{\rm MB}}

\newcommand{\tr}{{\rm Tr}}

\usepackage[usenames,dvipsnames]{color}
\definecolor{MyOrange}{rgb}{1.0,0.5,0}
\definecolor{MyPurple}{rgb}{0.5,0,1}

\begin{document}

\title{Non-equilibrium transport with self-consistent renormalised contacts 
for a single-molecule nanodevice with electron-vibron interaction}

\author{H. Ness}
\email{herve.ness@york.ac.uk}
\author{L. K. Dash}

\affiliation{Department of Physics, University of York, Heslington, York YO10 5DD,
UK}
\affiliation{European Theoretical Spectroscopy Facility (ETSF)}

\date{\today}

\begin{abstract}
  We present an application of a new formalism to treat the quantum
transport properties of fully interacting nanoscale junctions 
[Phys. Rev. B {\bf 84}, 235428 (2011)].
We consider a model single-molecule nanojunction in the presence of two kinds of 
electron-vibron interactions. In terms of electron density matrix, one interaction is diagonal 
in the central region and the second is off-diagonal in between the central region
and the left electrode. We use a non-equilibrium Green's function technique to calculate
the system's properties in a self-consistent manner. The interaction self-energies are
calculated at the Hartree-Fock level in the central region and at the Hartree level for the 
crossing interaction.
Our calculations are performed for different transport regimes ranging from the far off-resonance
to the quasi-resonant regime, and for a wide range of parameters.
They show that a non-equilibrium (i.e. bias dependent) static (i.e. energy independent) 
renormalisation is obtained for the nominal hopping matrix element between
the left electrode and the central region. Such a renormalisation is highly non-linear and
non-monotonic with the applied bias, however it always lead to a reduction of the current,
and also affects the resonances in the conductance.
Furthermore, we show that the relationship between the non-equilibrium 
charge susceptibility and dynamical conductance still holds even in the presence of crossing 
interaction. 

\end{abstract}
 

\maketitle

\section{Intro}
\label{sec:intro}


The theory of quantum transport in nano-scale devices has evolved rapidly over the past decade,
as advances in experimental techniques have made it possible to probe transport properties 
(at different temperatures) down to the single-molecule scale. 
Furthermore simultaneous measurement of charge and heat transport through single molecules is 
now also possible \cite{Widawsky:2012}. 
The development of accurate theoretical methods for the description of quantum transport at 
the single-molecule level is essential for continued progress in a number of areas including 
molecular electronics, spintronics, and thermoelectrics.

One of the longstanding problems of quantum charge transport is the establishment of a theoretical framework which allows for quantitatively accurate predictions of conductance from first principles. 
The need for methods going beyond the standard approach based on density functional theory combined 
with Landauer-like elastic 
scattering \cite{Hirose:1994,DiVentra:2000,Taylor:2001,Nardelli:2001,Brandbyge:2002,
  Gutierrez:2002,Frauenheim:2002,Xue:2003,Louis+Palacios:2003,Thygesen:2003,Garcia-suarez:2005}
 has been clear for a number of years. 
It is only recently that more advanced methods to treat electronic interaction have appeared, 
for example those based on the many-body $GW$ approximation \cite{Strange:2011,Rangel:2011,Darancet:2007}. Alternative frameworks to deal with the steady-state or time-dependent transport are given by 
many-body perturbation theory based on the non-equilibrium (NE) Green's function (GF) formalism:
in these approaches, the interactions and (initial) correlations are taken into account by using conserving approximations for the many-body self-energy \cite{Baym:1962,vonBarth:2005,vanLeeuwen:2006,Kita:2010,Tran:2008,Myohanen:2008,Myohanen:2010,
Perfetto:2010,Velicky:2010,
PuigvonFriesen:2009}.
 
Other kinds of interactions, e.g. electron-vibron coupling, also play an important role in 
single-molecule quantum transport. Inelastic tunneling spectroscopy constitutes an important 
basis for spectroscopy of molecular junctions, yielding insight into the vibrational modes 
and ultimately the atomic structure of the junction \cite{Arroyo:2010}.
There have been many theoretical investigations focusing on the
effect of electron-vibron coupling in
molecular and atomic scale wires \cite{Ness:1999, Ness:2001,
  Ness:2002a, Flensberg:2003, Mii:2003, Montgomery:2003b, Troisi:2003,
  Chen:2004, Lorente:2000, Frederiksen:2004, Galperin:2004,
  Galperin:2004b, Mitra:2004, Pecchia:2004, Pecchia:2004b,
  Chen_Z:2005, Paulsson:2005, Ryndyk:2005, Sergueev:2005, Viljas:2005,
  Yamamoto:2005, Cresti:2006, Kula:2006, Paulsson:2006, Ryndyk:2006,
  Troisi:2006b, Vega:2006, Caspary:2007, Frederiksen:2007,
  Galperin:2007, Ryndyk:2007, Schmidt:2007, Troisi:2007, Asai:2008,
  Benesch:2008, Paulsson:2008, Egger:2008, Monturet:2008,
  McEniry:2008, Ryndyk:2008, Schmidt:2008,
  Tsukada:2009,Loos:2009,Secker:2011,Hartle:2011a,Hartle:2011b}.
In all these studies, the interactions have always been considered to be present 
in the central region (i.e. the molecule) only, and the latter is connected to two non-interacting
terminals. Interactions are also assumed not to cross at the contracts between the central region
and the leads. When electronic interactions are present throughout the system, as within 
density-functional theory calculations, they are treated at the mean-field level and do not 
allow for any inelastic scattering events.
However, there are good reasons to believe that such approximations are only valid in a very 
limited number of practical cases. The interactions, in principle, exist throughout the entire
system.

In a recent paper we derived a general expression for the current in nano-scale junctions
with interaction present everywhere in the system \cite{Ness:2011}.
With such a formalism, we can calculate the transport properties in those systems
where the interaction is present everywhere.
The importance of extended interaction in nano-scale devices has also been addressed, 
for electron-electron interaction, in recently developed approaches such as Refs.~[\onlinecite{Strange:2011,Perfetto:2012}]. 

In the present paper, we also consider interactions existing beyond the central region.
We apply our recently developed formalism \cite{Ness:2011} for fully interacting systems
to a specific model of a single-molecule nanojunction.
We focuss on a model system in the presence of electron-vibron interaction 
within the molecule and between the molecule and one of the leads. We show how the interaction
crossing at one interface of the molecular nanojunctions affects the transport properties
by renormalising the coupling at the interface in a bias-dependent manner. We also study
the relationship between the \NE charge susceptibility \cite{Ness:2012} and
the dynamical conductance for the present model of interaction crossing at the contacts.

The paper is organised as follows:
In Sec. \ref{sec:transport}, we briefly recall the main result of our current expression
for fully interacting systems.
In Sec. \ref{sec:interac}, we present the model Hamiltonian for the system which include
two kinds of electron-vibron interaction, an Holstein-like Hamiltonian combined with a
Su-Schrieffer-Heeger-like Hamiltonian. In this section, we also describe how the corresponding
self-energies are calculated and the implications of such approximations on the current
expression at the left and right interfaces. 
In Sec. \ref{sec:res}, we show that our approximations are fully consistent with the constraint
of current conservation. Then the effects of the static non-equilibrium 
(i.e. energy-independent but bias-dependent) renormalisation of the coupling at the contact on 
both the current 
and the dynamical conductance are studied for a wide range of parameters.
We also show that the NE charge susceptibility is still related to the dynamical conductance
even in the presence of crossing interaction at the contact.
We finally conclude and discuss extensions of the present work in Sec. \ref{sec:ccl}.

\section{General theory for quantum transport}
\label{sec:transport}

We consider a two-terminal device, made of three regions left-central-right,
 in the steady-state regime.
In such a device, labelled $L-C-R$, the
interaction---which we specifically leave undefined
(e.g. electron-electron or electron-phonon)---is assumed to be
well described in terms of the single-particle self-energy $\Sigma^{\rm MB}$ and
spreads over the entire system. 

We use a compact notation for the Green's function $G$ and the self-energy $\Sigma$ 
matrix elements $M(\omega)$.  
They are annotated $M_C$ ($M_L$ or $M_R$) for the elements in the
central region $C$ (left $L$, right $R$ region respectively), and $M_{LC}$ (or $M_{CL}$)
and $M_{RC}$ (or $M_{CR}$) for the elements between region $C$ and region $L$ or $R$. 
There are no direct interactions between the two electrodes,
i.e. $\Sigma^{\rm MB}_{LR/RL}=0$.

In Refs.~[\onlinecite{Ness:2011,Ness:2012b}], we showed that for a finite applied bias $V$
the steady-state current $I_L(V)$ flowing through the left $LC$ interface is given by:
\begin{equation}
\label{eq:ILfinal}
\begin{split}
 I_L & = \frac{e}{\hbar} \int \frac{d\omega}{2\pi} \\
& \tr_{\{C\}}\left[ G^r_C \tilde{\Upsilon}^{L,l}_C + G^a_C
  (\tilde{\Upsilon}^{L,l}_C)^\dagger  + G^<_C(\tilde\Upsilon^L_C -
  (\tilde{\Upsilon}^L_C)^\dagger) \right] \\
+ & \tr_{\{L\}}\left[\Sigma^{\MB,>}_L G^<_L - \Sigma^{\MB,<}_L G^>_L \right]
\end{split}
\end{equation}
where the $\Upsilon_C$ quantities are 
\begin{equation}
\label{eq:Upsilons}
\begin{split}
\tilde\Upsilon^L_C(\omega)  & = \Sigma^a_{CL}(\omega)\ \tilde{g}^a_L(\omega)\ \Sigma^r_{LC}(\omega) , \\
(\tilde\Upsilon^L_C)^\dag & = \Sigma^a_{CL}\ \tilde{g}^r_L\ \Sigma^r_{LC} , \\
\tilde\Upsilon^{L,l}_C  & = \Sigma^<_{CL} \left( \tilde{g}^a_L - \tilde{g}^r_L \right) \Sigma^r_{LC}
+ \Sigma^r_{CL}\ \tilde{g}^<_L\ \Sigma^r_{LC} .
\end{split}
\end{equation}
By definition $\Sigma_{LC}(\omega)= V_{LC} + \Sigma^\MB_{LC}(\omega)$ (similarly for the $CL$
components) where $V_{LC/CL}$ are the nominal coupling matrix elements between the $L$ and $C$
regions.
$\tilde{g}^{x}_L(\omega)$ are the GF of the region $L$ renormalised by the interaction
{\em inside} that region, where $x=r,a,<$ stands for the retarded, advanced and lesser
GF components. For example, for the advanded and retarded components, we have
$(\tilde{g}^{r/a}_L(\omega))^{-1} = ({g}^{r/a}_L(\omega))^{-1} - \Sigma^{{\rm MB},r/a}_L(\omega)$
where all quantities are defined only in the subspace $L$.

The first line in the current equation Eq.~(\ref{eq:ILfinal}) corresponds to a 
generalisation of the Meir and Wingreen \cite{Meir:1992} result
to the cases for which the 
interactions are present in the three $L, C, R$ regions as well as in between the 
$L/C$ and $C/R$ regions. The second trace in Eq.~(\ref{eq:ILfinal}) corresponds to 
inelastic events induced by the interaction in the $L$ lead.
When a local detailed balance equation helds, this terms vanishes since locally one
has $\Sigma^{\MB,>} G^< = \Sigma^{\MB,<} G^>$.

Eq.~(\ref{eq:ILfinal}) bears some resemblance to the expression derived by Meir
and Wingreen \cite{Meir:1992} when written as:
\begin{equation}
\label{eq:IL_MeirWingreen_bis}
\begin{split}
I_L^{\rm MW} & = \frac{e}{\hbar} \int \frac{{\rm d}\omega}{2\pi} \\
& {\rm Tr}_{\{C\}} \left[ G^r_C \Sigma^{L,<}_C + G^a_C (\Sigma^{L,<}_C)^\dag  
+ G^<_C  (\Sigma^{L,a}_C - \Sigma^{L,r}_C) \right] .
\end{split}
\end{equation}
where we use the definitions
$\Sigma^{L,<}_C = - (\Sigma^{L,<}_C)^\dag = V_{CL}\ {g}^<_L\ V_{LC} = {\rm i} f_L \Gamma_L$
and
$\Sigma^{L,a}_C - \Sigma^{L,r}_C = V_{CL} ({g}^a_L-g^r_L) V_{LC} = {\rm i}\Gamma_L$. 
Hence $I_L^{\rm MW}$ becomes

Hence one can see by comparing Eq.~(\ref{eq:ILfinal}) and Eq.~(\ref{eq:IL_MeirWingreen_bis})
that the quantities $\tilde\Upsilon_{LC}$ ($\tilde\Upsilon^\dag_{LC}$)
and $\tilde\Upsilon^l_{LC}$ 
are playing the role of the $L$ lead self-energy  
$\Sigma^a_L$ ($\Sigma^r_L$) and $\Sigma^<_L$ respectively
when the interactions cross at the $LC$ interface.
In Meir and Wingreen model, the leads are non-interacting, hence the second trace 
$\tr_{\{L\}}\left[...\right]$ in Eq.~(\ref{eq:ILfinal}) does not exist.

\section{Model for the interaction}
\label{sec:interac}

\subsection{Hamiltonians}
\label{sec:Hamiltonian}

We consider a single-molecule junction in the presence of electron-vibron interaction 
inside the central region and crossing at the contacts. 
Using a model system to reduce these calculations to a tractable size, we
concentrate on a single molecular level coupled to a single
vibrational mode.  A full description of our methodology, for the interaction inside
the region $C$, is provided in Refs.~[\onlinecite{Dash:2010,Ness:2010,Dash:2011}].
Furthermore, we consider that the electron-vibron interaction exist also at one
contact (the left $L$ electrode for instance).
This model typically corresponds to an experiment for a molecule chemisorbed onto a surface (the
left electrode) with a tunneling barrier to the right $R$ lead.

In the following model, we consider two kinds of electron-vibron coupling: 
a local coupling in the sense of an Holstein-like coupling of the electron charge density with a 
internal degree of freedom of vibration inside the central region, 
and an off-diagonal coupling in the sense of a Su-Schrieffer-Heeger-like coupling \cite{Heeger:1988,Ness:2001} to another vibration mode involving the hopping of an electron between 
the central $C$ region and the $L$ electrode.

The Hamiltonian for the region $C$ is
\begin{equation}
\label{eq:H_central}
\begin{split}
  H_C 
  = \varepsilon_0 d^\dagger d + \hbar \omega_0 a^\dagger a +
  \gamma_0 (a^\dagger + a) d^\dagger d,
\end{split}
\end{equation}
where $d^\dagger$ ($d$) creates (annihilates) an
electron in the molecular level $\varepsilon_0$. The electron charge
density in the molecular level is coupled to the vibration mode 
of energy $\omega_0$ via the coupling constant $\gamma_0$, and
$a^\dagger$ ($a$) creates (annihilates) a vibration quantum 
in the vibron mode $\omega_0$.
The central region $C$ is nominally connected to two (left and right) one-dimensional
tight-binding chains via the hopping integral $t_{0L}$ and
$t_{0R}$. The corresponding electrode $\alpha=L,R$ self-energy is
$\Sigma^r_\alpha(\omega)=t_{0\alpha}^2/\beta_\alpha \exp^{{\rm i} k_\alpha(\omega)}$
with the dispersion relation 
$\omega=\varepsilon_\alpha+2\beta_\alpha \cos(k_\alpha(\omega))$ where
$\varepsilon_\alpha$ and $\beta_\alpha$ are the tight-binding on-site and off-diagonal
elements of the electrode chains.

To describe the electron-vibron interaction existing at the left contact, we consider
that the hopping integral $t_{0L}$ is actually dependent on some generalised coordinate $X$.
The latter represents either the displacement of the centre-of-mass of the molecule or of
some chemical group at the end of the molecule link to the $L$ electrode. 
At the lowest order, the matrix element can be linearised as
$t_{0L}(X)= t_{0L}+t'_{0L} X$.
Hence the hopping of an electron from the $C$ region to the $L$ region (and {\em vice versa})
is coupled to a vibration mode (of energy $\omega_A$) via the coupling constant $\gamma_A$
(itself related to $t'_{0L}$).
The corresponding Hamiltonian is given by 
\begin{equation}
\label{eq:Vcrossing}
H_{LC} = \gamma_A (b^\dagger + b)(c^\dagger_L d + d^\dagger c_L) + \omega_A b^\dagger b , 
\end{equation}
where $b^\dagger$ ($b$) creates (annihilates) a vibration quantum in the vibron mode $\omega_A$,
the generalised coordinate is $X=\sqrt{{\hbar}/{(2m_A\omega_A)}}(b^\dagger + b)$,
and $c^\dagger_L$ ($c_L$) creates (annihilates) an
electron in the level $\varepsilon_L$ of the $L$ electrode.

The Hamiltonians Eq.~(\ref{eq:H_central}) and Eq.~(\ref{eq:Vcrossing}) are used to calculate 
the corresponding electron self-energies at different orders of the interaction $\gamma_0$ and 
$\gamma_A$ using conventional \NE diagrammatics techniques \cite{Dash:2010,Dash:2011}.

Furthermore, at equilibrium, the whole system has a single and well-defined Fermi 
level $\mu^{\rm eq}$. 
A finite bias $V$, applied across the junction, lifts the Fermi levels as
$\mu_{L,R}=\mu^{\rm eq}+\eta_{L,R} eV$.  The fraction of potential drop \cite{Datta:1997} 
at the left contact is $\eta_L$ and $\eta_R=\eta_L-1$ at the right contact, 
with $\mu_L-\mu_R=eV$ and $\eta_L \in [0,1]$.

\subsection{Self-energies for the interactions}
\label{sec:SE}

The electron-vibron \SEs in the central region $C$ are calculated within the Born 
approximation.
The details of the calculations are reported elsewhere \cite{Dash:2010,Dash:2011} so
we briefly recall the different expressions for the \SEs 
$\Sigma^{\MB,x}_C=\Sigma^{H,x}_C + \Sigma^{F,x}_C$ with
\begin{equation}
\label{eq:SigmaC_Hartree}
\begin{split}
  \Sigma^{H,r}_C =  \Sigma^{H,a}_C =  2 \frac{\gamma_0^2}{\omega_0} 
\int \frac{d\omega^\prime}{2\pi} i G^<_C(\omega^\prime) 
 =  - 2 \frac{\gamma_0^2}{\omega_0} \langle n_C \rangle \ ,
\end{split}
\end{equation}
with $\langle n_C \rangle = - i \int {d\omega}/{2\pi}\  G^<_C(\omega)$ 
and
\begin{equation}
\label{eq:SigmaClessgrt_Fock}
\Sigma^{F,\lessgtr}_C(\omega)  =  i \gamma_0^2 
\int \frac{d u}{2\pi} \
    D_0^\lessgtr(u) \
    G^\lessgtr_C(\omega - u) \ ,
\end{equation}
and
\begin{equation}
\label{eq:SigmaCr_Fock}
\begin{split}
\Sigma^{F,r}_C(\omega)  = i \gamma_0^2 
      \int \frac{d u}{2\pi} & D_0^r(\omega - u)
      \left( G^<_C(u) + G^r_C(u) \right)  \\ 
+ & D_0^<(\omega -u) G^r_C(u) \ ,
\end{split}
\end{equation}
with the usual definitions for the bare vibron GF $D_0^x$:
\begin{equation}
\label{eq:D0}
  \begin{split}
    D_0^\lessgtr(\omega) & = -2\pi i \left[ \langle N_0 \rangle 
\delta(\omega \mp \omega_0) + \langle {N_0} \rangle + 1) \delta(\omega \pm \omega_0) \right] \\
    D_0^r(\omega) & = \frac{1}{\omega - \omega_0 +i 0^+} 
    - \frac{1}{\omega + \omega_0 +i 0^+} \ , 
  \end{split}
\end{equation}
where $\langle {N_0} \rangle$ is the averaged number of excitations in
the vibration mode of frequency $\omega_0$ given by the Bose-Einstein 
distribution at temperature $T_{\rm vib}$. In the following, we work in
the limit of low temperature for which  $\expect{N_0}=0$.

As a first application of our transport formalism for crossing interactions,
we consider a mean-field approximation for the electron-vibron coupling at
the $LC$ interface. 
This leads to the Hartree-like expressions for the many-body self-energies 
at the $LC$ interface:  
\begin{equation}
\label{eq:Vcrossing_Hartree}
\Sigma^{\MB,r/a}_{LC}=- 2\frac{\gamma_A^2}{\omega_A} \langle n_{LC}\rangle ,
\end{equation}
where
\begin{equation}
\label{eq:nLC}
\langle n_{LC}\rangle = - i \int\frac{{\rm d}\omega}{2\pi}\ G^<_{LC}(\omega) .
\end{equation}
Similarly the self-energy $\Sigma^{\MB,r/a}_{CL}$ is obtained from 
$\langle n_{CL}\rangle = -i \int{\rm d}\omega / {2\pi}\ G^<_{CL}(\omega)$.

One can see that the interaction crossing at the $LC$ interface induces a static 
(however bias-dependent) renormalisation of the nominal coupling $V_{CL}=V_{LC}=t_{0L}$ 
between the $L$ and $C$ regions. 
This \NE renormalisation will induce, amongst other effects, a bias-dependent modification of the broadening of the spectral features of the $C$ region.
Since the renormalisation is static at the Hartree level, we use a ``potential'' notation to represent 
the normalised coupling: $\Sigma^{r/a}_{LC}\equiv\tilde V_{LC}$ 
with
\begin{equation}
\label{eq:VLC_renorm}
\tilde V_{LC} = V_{LC} - 2\frac{\gamma_A^2}{\omega_A} \langle n_{LC}\rangle,
\end{equation}
and similarly for $\tilde V_{CL}$.

The static renormalisation of the nominal coupling $t_{0L}$ is driven by the 
ratio ${\gamma_A^2}/{\omega_A}$. In the following
numerical applications, we consider small to larger renormalisation effects, for which
the ratio ${\gamma_A^2}/({\omega_A t_{0L}})$ is ranging from $\sim 0.1$ to $\sim 5.0$.
One should note, however, that in all the calculations we have performed, the density
matrix element $\langle n_{LC}\rangle$ is always of the order of $10^{-2}$. 
Therefore the renormalisation effects are always smaller than the nominal coupling
$t_{0L}$ itself.

\begin{widetext}

In order to get the renormalised couplings $\tilde V_{LC/CL}$, we need the off-diagonal 
elements $G^<_{LC}$ and $G^<_{CL}$. 
The closed expression for the $G^<_{LC}$ GF matrix element is obtained from the corresponding 
Dyson equation $G^<_{LC}=[g\Sigma G]^<_{LC}$. 
After formal manipulation, we find that
\begin{equation}
\label{eq:GlesserLC}
G^<_{LC}(\omega) = 
g^<_L \tilde V_{LC} G^a_C +
G^r_{LC} 
\left(
\tilde V_{CL} g^<_L \tilde V_{LC} + \Sigma^<_C + V_{CR} g^<_R V_{RC}
\right)
G^a_C , 
\end{equation}
with $G^a_C$ is the renormalised advanced GF of the central region
\begin{equation}
\label{eq:GaC}
G^a_C(\omega) = 
\left[
[g^a_C]^{-1} -\Sigma^a_C - V_{CR} g^a_R V_{RC} - \tilde V_{CL} g^a_L \tilde V_{LC}
\right]^{-1} ,
\end{equation}
and the off-diagonal element $G^r_{LC}$ is given by
\begin{equation}
\label{eq:GrLC}
G^r_{LC}(\omega) = G^r_L \tilde V_{LC} 
\left[
[\tilde g^r_C]^{-1} - V_{CR} g^r_R V_{RC}
\right]^{-1} ,
\end{equation}
where $\tilde g^r_C$ is the renormalised GF of the disconnected C region
$[\tilde g^r_C]^{-1}= [g^r_C]^{-1} - \Sigma^r_C$.
After further manipulation, we get the following expression for $G^r_{LC}$
by using the notation $Y_C^{R,r} = V_{CR} g^r_R V_{RC}$ for the non-interacting
$R$ lead self-energy:
\begin{equation}
\label{eq:GrLC_bis}
G^r_{LC}(\omega) =
\left[
[g^r_L]^{-1} - 
\tilde V_{LC}
\left[ [\tilde g^r_C]^{-1} - Y_C^{R,r} \right]^{-1}
\tilde V_{CL}
\right]^{-1} 
\tilde V_{LC}
\left[
[\tilde g^r_C]^{-1} - Y_C^{R,r}
\right]^{-1} .
\end{equation}

Similarily we can calculate the off-diagonal element $G^<_{CL}$ from the corresponding Dyson
equations $G^<_{CL}=[g\Sigma G]^<_{CL}$. We find
\begin{equation}
\label{eq:GlesserCL}
G^<_{CL}(\omega) = 
G^r_C \tilde V_{CL} g^<_L +
G^r_C 
\left(
\tilde V_{CL} g^<_L \tilde V_{LC} + \Sigma^<_C + V_{CR} g^<_R V_{RC}
\right)
G^a_{CL} , 
\end{equation}
where $G^r_C(\omega)$ is the retarded version of Eq.~(\ref{eq:GaC})
and 
\begin{equation}
\label{eq:GaCL}
G^a_{CL}(\omega) = 
\left[
[\tilde g^a_C]^{-1} - Y_C^{R,a} \right]^{-1} 
\tilde V_{CL} G^a_L  .
\end{equation}

As expected from the definition of the different GFs, we can see that indeed 
$(G^<_{CL}(\omega))^* = - G^<_{LC}(\omega)$
and
$(G^a_{CL}(\omega))^* =   G^r_{LC}(\omega)$.

\end{widetext}

\subsection{Calculations}
\label{sec:calc_method}

Calculations are performed in a self-consistent manner. There are different ways to calculate the
GF and the \SEs in a self-consistent way: in the present work, we first perform self-consistent
calculations for the central region $C$. 
The new hopping matrix elements between the $C$ and $L$ regions given by Eq.~(\ref{eq:VLC_renorm}) 
are then updated via the use of Eqs.~(\ref{eq:GlesserCL}) and (\ref{eq:GlesserLC}). 
The calculations are then re-iterated until full self-consistency for all self-energies and GF 
in the region $C$ and at the left contact $\Sigma^{\MB,r/a}_{LC/CL}$ is obtained.

For the model of crossing interaction we are considering here, further simplications can be
introduced in the calculation of the current.
Since there is no other interaction inside the $L$ and $R$ regions, the
current expression is given only by the first line of Eq.(\ref{eq:ILfinal}).
Furthermore, there are no lesser and greater components for the self-energy $\Sigma^{\MB}_{LC}$ 
at the mean-field level. 
The current expression Eq.~(\ref{eq:ILfinal}) reduces to:
\begin{equation}
\label{eq:IL_Hartree}
\begin{split}
 I_L & = \frac{e}{\hbar} \int \frac{d\omega}{2\pi} \\
& \tr_{\{C\}}\left[ G^r_C \tilde{\Upsilon}^{L,l}_C + G^a_C
  (\tilde{\Upsilon}^{L,l}_C)^\dagger  + G^<_C(\tilde\Upsilon^L_C -
  (\tilde{\Upsilon}^L_C)^\dagger) \right] \ ,
\end{split}
\end{equation}
with the following simplified expressions for the $\Upsilon_{C}$ quantities 
\begin{equation}
\label{eq:Hartreenorm}
\begin{split}
\tilde\Upsilon^L_C           & = \tilde V_{CL}\ {g}^a_L\ \tilde V_{LC} , \\ 
(\tilde\Upsilon^L_C)^\dag      & = \tilde V_{CL}\ {g}^r_L\ \tilde V_{LC}, \\
\tilde\Upsilon^{L,l}_{C}         & = \tilde V_{CL}\ {g}^<_L\ \tilde V_{LC}, \\
(\tilde\Upsilon^{L,l}_{C})^\dag  & = - \tilde V_{CL}\ {g}^<_L\ \tilde V_{LC}, \\
\end{split}
\end{equation}
since $\Sigma^{r/a}_{LC}=\tilde V_{LC}$ and $\Sigma^\lessgtr_{LC}=0$.

Furthermore, the current at the $CR$ interface has a conventional Meir and Wingreen 
expression, since there are no crossing interactions:
\begin{equation}
\label{eq:IR_MeirWingreen}
\begin{split}
I_R^{\rm MW} & = - \frac{e}{\hbar} \int \frac{{\rm d}\omega}{2\pi} \\
& {\rm Tr}_{\{C\}} \left[ G^r_C \Sigma^{R,<}_C + G^a_C (\Sigma^{R,<}_C)^\dag  
+ G^<_C  (\Sigma^{R,a}_C - \Sigma^{R,r}_C) \right] \ ,
\end{split}
\end{equation}
where we use the conventional definition for the $R$ lead self-energies:
$\Sigma^{R,x}_C(\omega)= Y_C^{R,r}(\omega) = V_{CR} g^x_R(\omega) V_{RC}$
with $V_{CR}=V_{RC}=t_{0R}$.

The main differences between the expression for the left and right currents $I_L$
and $I_R^{\rm MW}$ arise from the non-equilibrium static renormalisation of the
coupling between the central region $C$ and the $L$ lead given in Eq.~(\ref{eq:VLC_renorm}).

\begin{figure}
  \includegraphics[width=72mm]{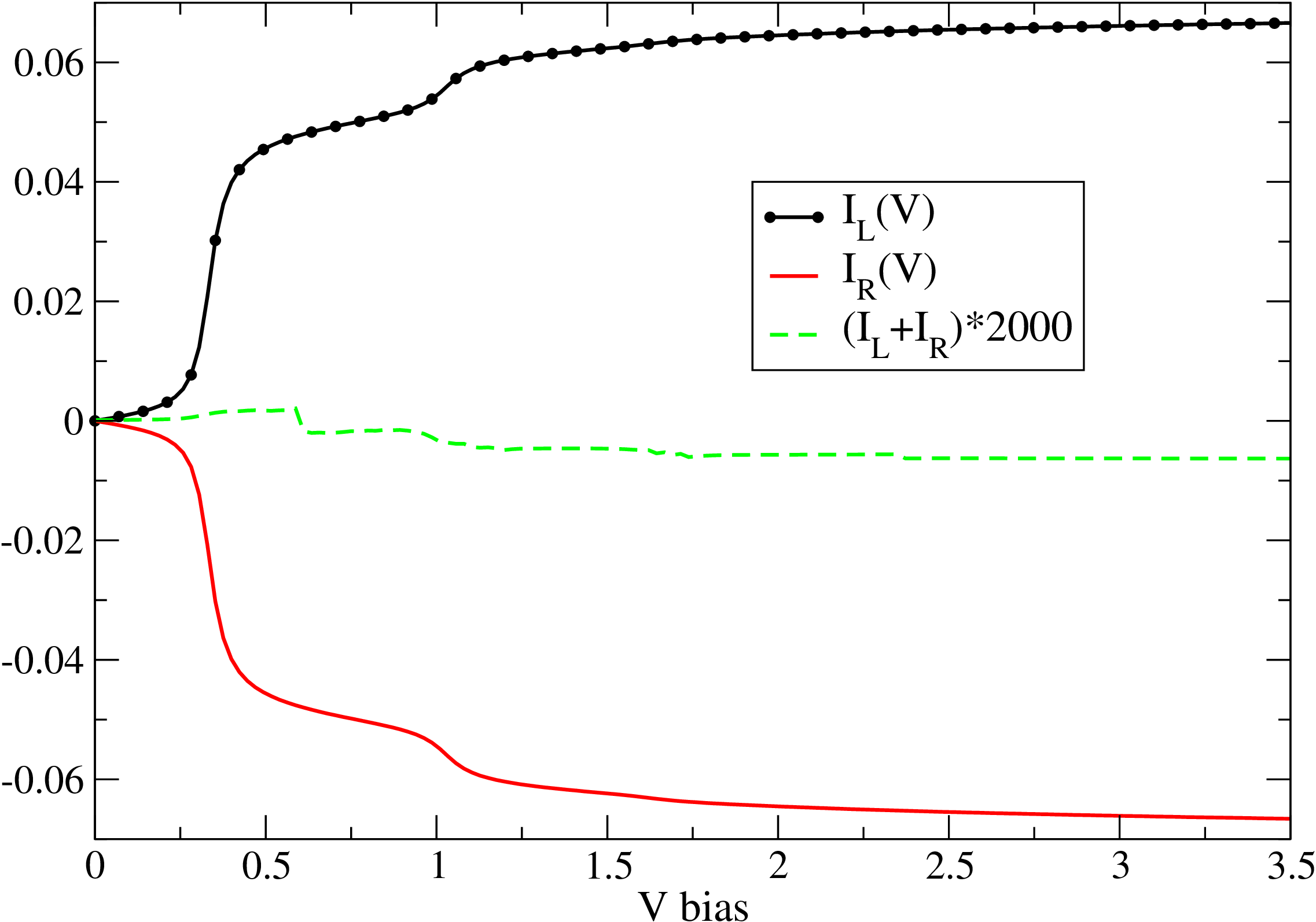}
  \caption{(Color online) Current at the left and right interfaces $I_L(V)$ (black line with dots)
  and $I_R(V)$ (red line) given respectively by Eq.~(\ref{eq:IL_Hartree}) and 
  Eq.~(\ref{eq:IR_MeirWingreen}). The current is fully conserved as can be seen by
  the green dashed line which shows $I_L+I_R$ magnified by a factor of 2000. 
  The parameters are $\varepsilon_0=0.3$, $\omega_0=0.30$, $\gamma_0=0.21$, $t_{0R}=t_{0L}=0.15$, 
  $\gamma_A=0.11$, $\omega_A=0.10$,
  $\beta_\alpha=2.0, \epsilon_\alpha=0.0$.}
  \label{fig:ex_Iconserv}
\end{figure}

\begin{figure}
  \includegraphics[width=72mm]{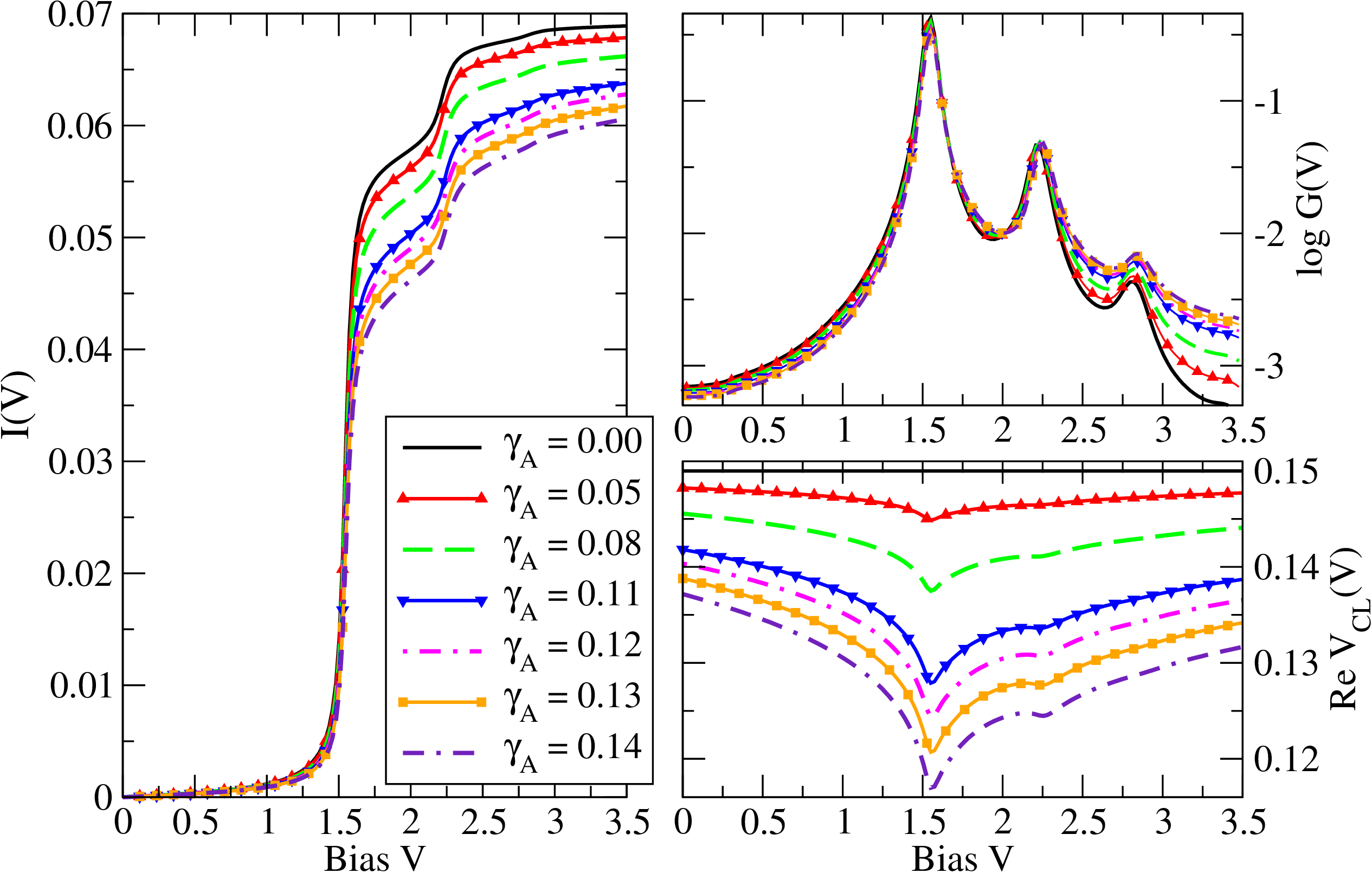}
  \caption{(Color online) Far off-resonant transport regime with $\varepsilon_0=0.9$.
  Current $I(V)$ (left panel), dynamical conductance $G(V)=dI/dV$ in logarithmic scale
  (top-right panel) and the real part of non-equilibrium renormalized static coupling 
  $V_{CL}(V)$ 
  (bottom-right panel) for different values of the coupling $\gamma_A$ strength at the
  left interface between $0.0$ and $0.14$, corresponding
  to the follow ratio: 
  ${\gamma_A^2}/(\omega_A t_{0L})=0.0, 0.167, 0.427, 0.807, 0.96, 1.127, 4.667$.
  The NE renormalization of $\tilde V_{CL}(V)$ corresponds to a non-monotonic reduction of
  the nominal hopping integral $t_{0L}$ with applied bias, which induces a diminution
  of the current, and modifications of the widths of the conductance peaks accompanied
  with a slight shift of the conductance peaks at high bias. See text for a complete
  discussion.
  The other parameters are $\omega_0=0.30$, $\gamma_0=0.21$, $t_{0R}=t_{0L}=0.15$, $\gamma_A=0.10$,
  $\beta_\alpha=2.0, \epsilon_\alpha=0.0$.}
  \label{fig:eps090}
\end{figure}

\section{Results}
\label{sec:res}

We have perfomed calculations for many different values of the parameters
in the Hamiltonians. We present below the most characteristic results for
different transport regimes and for different coupling strengths $\gamma_A$,
while the interaction in the region $C$ is taken to be in the intermediate 
coupling regime $\gamma_0/\omega_0 = 0.7$. 
The nominal couplings between the central region and the electrodes $t_{0L,R}$, 
before NE renormalisation, are not too large, so that we can discrimate clearly 
between the different vibron side-band peaks in the spectral functions.
The values chosen for the parameters are typical values for realistic molecular 
junctions \cite{Dash:2012,Dash:2011}.

The different transport regimes considered bellow are called
the far off-resonant, the off-resonant, the intermediate regime
and the quasi-resonant regime.
They correspond to different position of the molecular level $\varepsilon_0$
with respect to the Fermi level $\mu^{\rm eq}$ at equilibrium.
With the static renormalised molecular level 
$\tilde\varepsilon_0=\varepsilon_0 - \gamma_0^2/\omega_0$, the far off-resonant
regime corresponding to $\tilde\varepsilon_0 - \mu^{\rm eq} \gg 0$ and
$\gg$ several $\omega_0$, the off-resonant regime corresponds to 
$\tilde\varepsilon_0 - \mu^{\rm eq} \gg 0$ and
$\sim$ one or two $\omega_0$, 
the intermediate regime corresponds to $\tilde\varepsilon_0 - \mu^{\rm eq} \sim \omega_0$
and the quasi-resonant
regime corresponds to $\tilde\varepsilon_0 - \mu^{\rm eq} < \omega_0$.

In the following the current is given in units of charge per time, the conductance
in unit of quantum of conductance $G_0=2e^2/\hbar$ and the bias $V$ and the 
normalised coupling $\tilde V_{CL}$ in natural units of energy where $e=1$ and $\hbar=1$.

\subsection{Current conservation}
\label{sec:Iconserv}

One of the most important physical conditions that our formalism needs to 
fulfil is the constraint of current conservation. We use conserving approximations
to calculate the interaction self-energies in the central region $C$ and for the
crossing interaction at the left interface. 
However, since there is no interaction crossing at the $CR$ interface while the
interaction \emph{is} crossing at the $LC$ interface, we have to check that the
current given by Eq.~(\ref{eq:IL_Hartree}) for $I_L(V)$ is equal to the Meir and Wingreen
current given by Eq.~(\ref{eq:IR_MeirWingreen}) for $I_R(V)$, i.e. $I_L+I_R=0$.
Figure \ref{fig:ex_Iconserv} shows that the condition of current conservation is
indeed fulfilled, as expected. 
We have carefully checked that the current is conserved for all the calculations 
presented in the present paper, i.e. that 
$\vert I_L+I_R \vert / \vert I_L \vert \sim \vert I_L+I_R \vert / \vert I_R \vert < 10^{-5}$.

\subsection{Static non-equilibrium renormalisation}
\label{sec:staticNErenorm}

\begin{figure}
  \includegraphics[width=72mm]{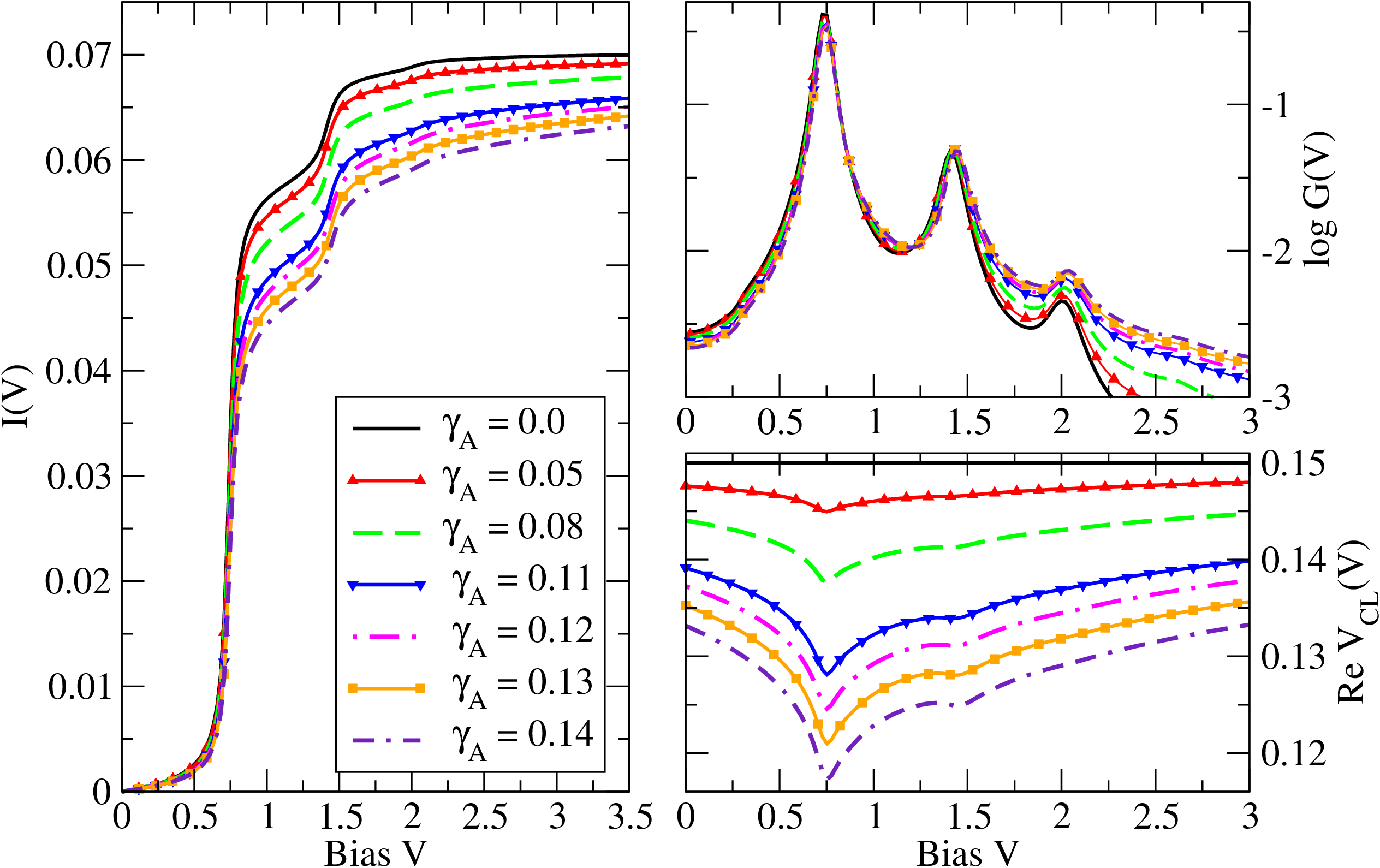}
  \caption{(Color online) Intermediate off-resonant transport regime with $\varepsilon_0=0.5$.
  The other parameters are the same as in figure \ref{fig:eps090}. 
  See text for a complete
  discussion.}
  \label{fig:eps050}
\end{figure}

\begin{figure}
  \includegraphics[width=72mm]{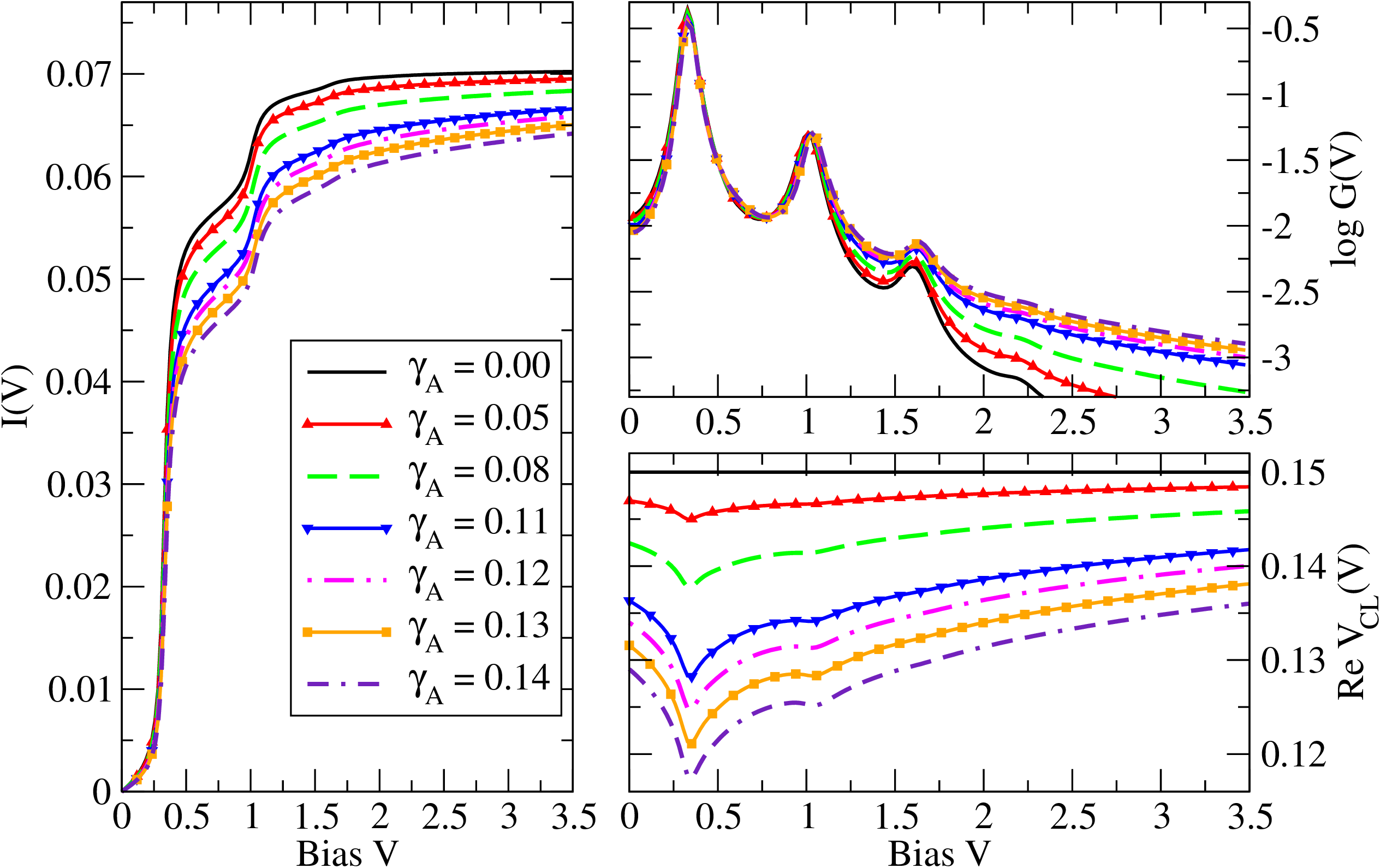}
  \caption{(Color online) Intermediate off-resonant/resonant regime with $\varepsilon_0=0.3$.
  The other parameters are the same as in figure \ref{fig:eps090}. See text for a complete
  discussion.}
  \label{fig:eps030}
\end{figure}

\begin{figure}
  \includegraphics[width=72mm]{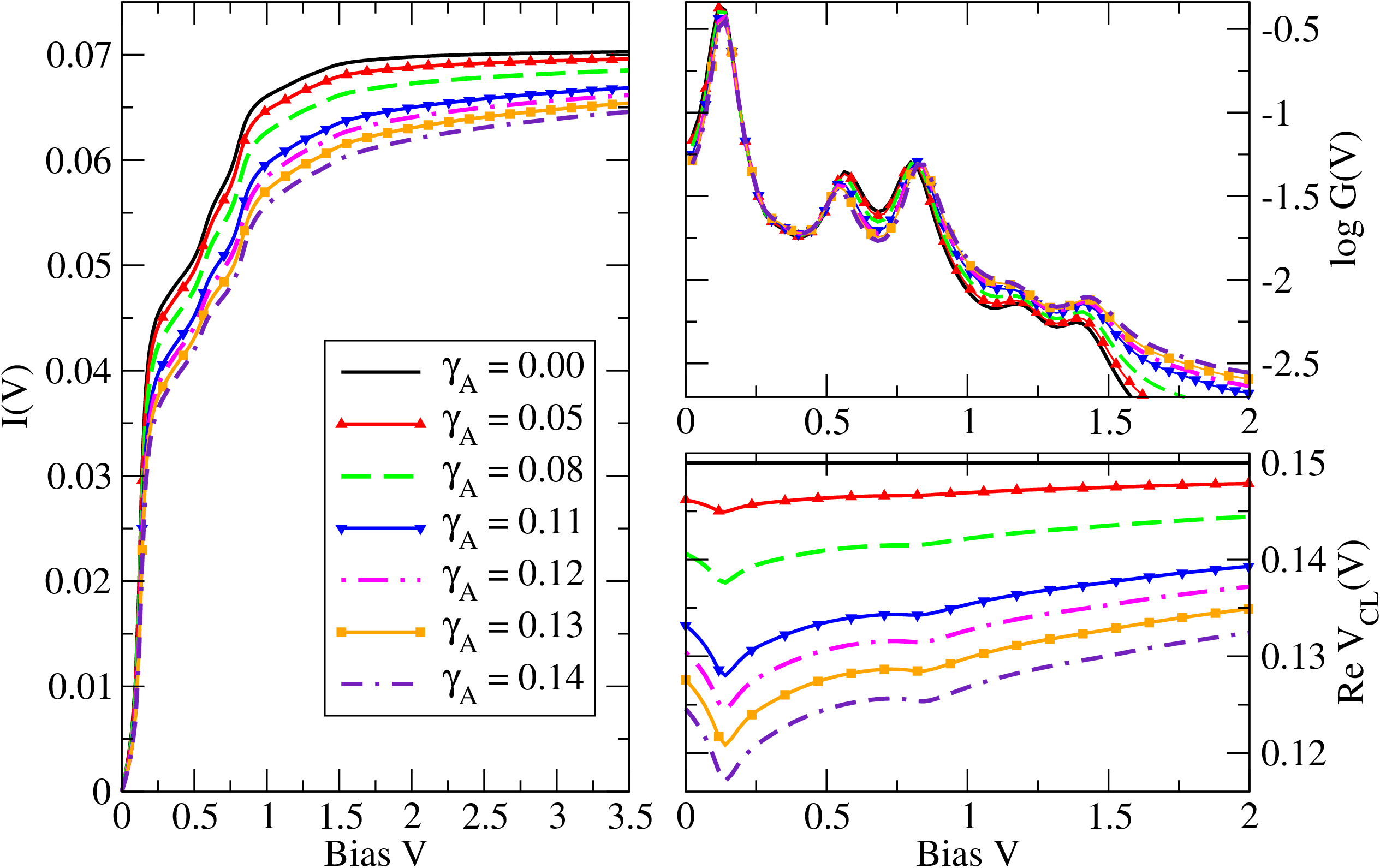}
  \caption{(Color online) Quasi-resonant transport regime with $\varepsilon_0=0.2$.
  Same parameters as in figure \ref{fig:eps090}. See text for a complete
  discussion.}
  \label{fig:eps020}
\end{figure}

In Figures \ref{fig:eps090} to \ref{fig:eps020}, we show the dependence of the
current $I(V)$, of the dynamical conductance $G(V)=dI/dV$ and of the renormalised
coupling $\tilde V_{CL}(V)$ on the applied bias $V$, for different values of the
interaction strength $\gamma_A$ at the left contact.
We consider different transport regimes ranging from the far off-resonant regime
(Fig.~\ref{fig:eps090}), 
the off-resonant regime (Fig.~\ref{fig:eps050}), 
the intermediate regime (Fig.~\ref{fig:eps030}) and finally  
the quasi-resonant regime (Fig.~\ref{fig:eps020}).

The NE renormalisation of the coupling at the left contact
given by Eq.~(\ref{eq:VLC_renorm}) corresponds to an effective decrease of the
hopping integral leading to a decrease of the current for increasing values of
$\gamma_A$, as can be seen in the left panels of figs.~\ref{fig:eps090} to \ref{fig:eps020}.
The real part of $\tilde V_{CL}(V)$ shows a non-monotonic behaviour
with the applied bias and presents features (local dips) at applied biases corresponding
to peaks in the dynamical conductance (bottom right panels of figs.~\ref{fig:eps090} to \ref{fig:eps020}).
Therefore, the interaction crossing at the left contact not only decreases the value
of the current but also affects the width of the peaks in the conductance, as can be seen
in the top right panels of figs.~\ref{fig:eps090} to \ref{fig:eps020}.

In all our calculations, it appears that each conductance peak now has an asymmetric
shape, i.e. a different broadening on each side of the peak, which is due to the
non-monotonic and asymmetric behaviour of $\tilde V_{CL}$ versus applied bias.
The detailed understanding of such a behaviour is rather complex. 
However, one can obtain a qualitative understanding of our results by considering the 
following analysis.

We can use the current expressions for $I_L$ and $I_R$ given
by Eq.~(\ref{eq:IL_Hartree}) and Eq.~(\ref{eq:IR_MeirWingreen}) respectively
to obtain
a symmetrized current $I=(I_L-I_R)/2$ expression within a good approximation \cite{Meir:1992} :
\begin{equation}
\label{eq:Isym}
 I = \frac{e}{\hbar} \int \frac{d\omega}{2\pi} \
 (f_L-f_R)\ \frac{\tilde\Gamma_C^L \Gamma_C^R}{\tilde\Gamma_C^L + \Gamma_C^R}\
\pi A_C(\omega) \ ,
\end{equation}
where $\pi A_C(\omega) = - \Im m G_C^r(\omega)$ and
\begin{equation}
\label{eq:G^r_C}
G_C^r = \left[
\omega - \varepsilon_0 - \Sigma_C^{H+F,r} - \tilde\Sigma_C^{L,r} 
- \Sigma_C^{R,r} \right]^{-1} \ ,
\end{equation}
with the self-energy in the central region $\Sigma_C^{H+F,r}(\omega)$
given by the sum of Eq.~(\ref{eq:SigmaC_Hartree}) and Eq.~(\ref{eq:SigmaCr_Fock}); 
and we recall that 
\begin{equation}
\label{eq:Sigma_leads} 
\begin{split}
\tilde\Sigma_C^{L,r}(\omega) =   \tilde V_{CL}\ {g}^r_L(\omega)\ \tilde V_{LC} =
\Re e \Sigma_C^{L,r} - i \tilde\Gamma_C^L/2 , \\
      \Sigma_C^{R,r}(\omega) =          V_{CR}\ {g}^r_R(\omega)\        V_{RC} =
\Re e \Sigma_C^{R,r} - i       \Gamma_C^R/2 .
\end{split}
\end{equation}

In the off-resonant regime and for small bias $V<\omega_0$, one can consider that
$\Im m \Sigma_C^{H+F,r} \sim 0$ (see Ref.~[\onlinecite{Ness:2011}]), hence
\begin{equation}
\label{eq:Isym_bis}
 I \propto \int \frac{d\omega}{2\pi} \
 \frac{(f_L-f_R)\ \tilde\Gamma_C^L \Gamma_C^R}
{(\omega - \varepsilon_0 - \Re e \Sigma_C^{\rm tot})^2+(\tilde\Gamma_C^L+\Gamma_C^R)^2/4} \ .
\end{equation}

\begin{figure}
  \includegraphics[width=72mm]{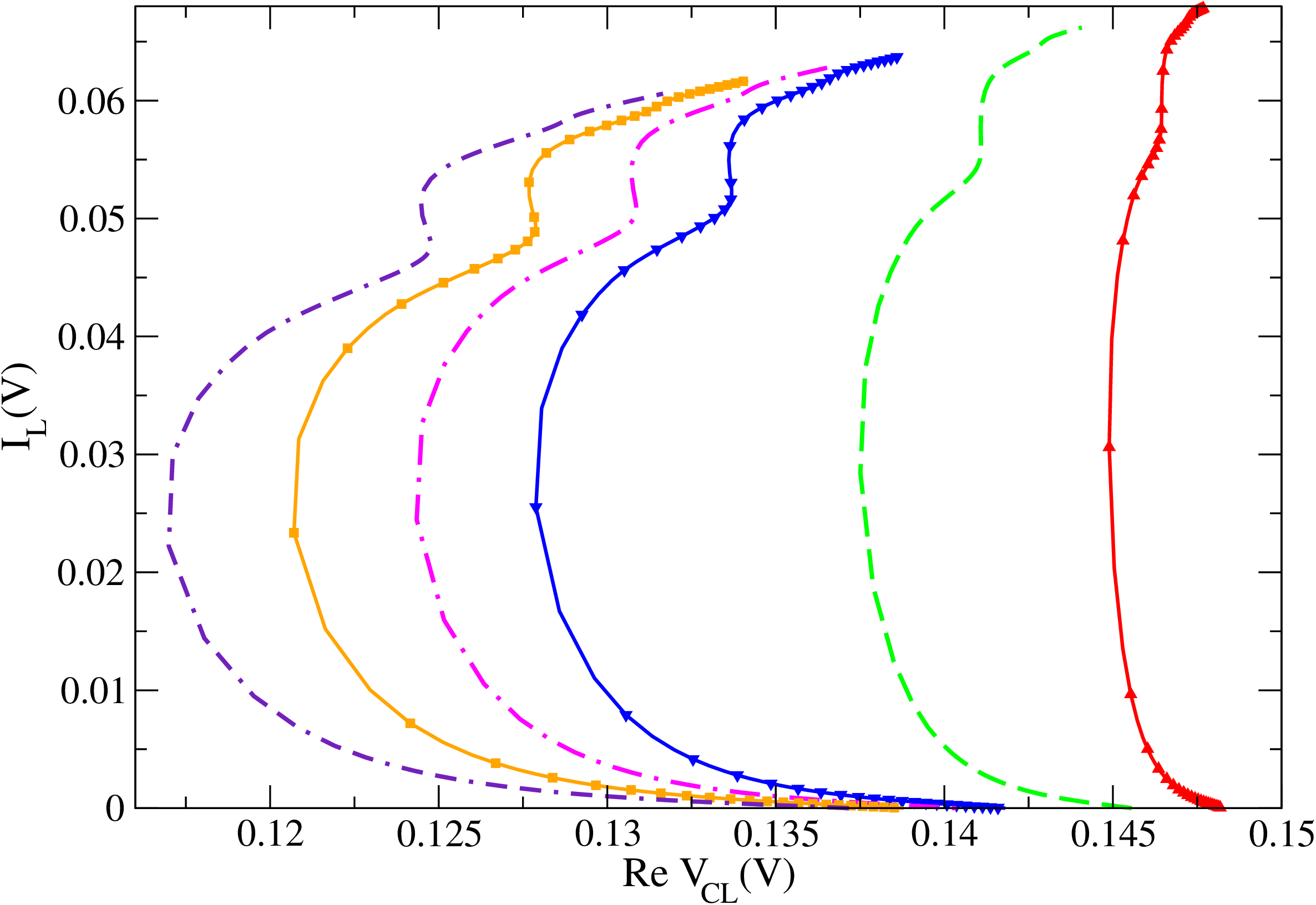}
  \caption{(Color online) Parametric curve $I(V) - \tilde V_{CL}(V)$ for the far off-resonant
  transport regime shown in Fig. \ref{fig:eps090} (using the same legend for the different
  $\gamma_A$ values). 
  The starting points of the parametric
  curves, i.e. $V=0$, are at the bottom of the graph (zero current values), and increasing
  $V$ corresponds to increasing values of the current. At low bias ($I_L < 0.01$),  
  $I_L$ is quadratic in $\tilde V_{CL}$.}
  \label{fig:parametricI_VCL}
\end{figure}

We can now use Eq.~(\ref{eq:Isym_bis}) to understand the behaviour of the current. In Figure
\ref{fig:parametricI_VCL} we show the parametric curves $I(V) - \tilde V_{CL}(V)$ obtained for 
the far off-resonant transport regime (shown in Fig.~\ref{fig:eps090}). Figure
\ref{fig:parametricI_VCL} shows a complex dependence of the current versus the NE renormalised
coupling $\tilde V_{CL}$. 
However at low bias, one can consider that $\tilde\Gamma_C^L$ and $\Gamma_C^R$
are independent of $\omega$ and hence take such quantities out of the integral in 
Eq.~(\ref{eq:Isym_bis}). Therefore we have  $I \propto \tilde\Gamma_C^L$ and the current 
is quadratic in $\tilde V_{CL}$ since $\tilde\Gamma_C^L \propto (\tilde V_{CL})^2$. 
Such a dependence can be clearly seen in Figure \ref{fig:parametricI_VCL} for the low bias 
regime where $I_L < 0.01$.

At larger bias, one can no longer neglect the $\omega$ dependence of $\tilde\Gamma_C^L$ 
(and $\Gamma_C^R$) and more importantly the effects of the interaction in the central region, i.e. 
$\Im m \Sigma_C^{H+F,r} \ne 0$. Hence the quadratic dependence of $I$ on $V_{CL}$ is lost.

\begin{figure}
  \includegraphics[width=72mm]{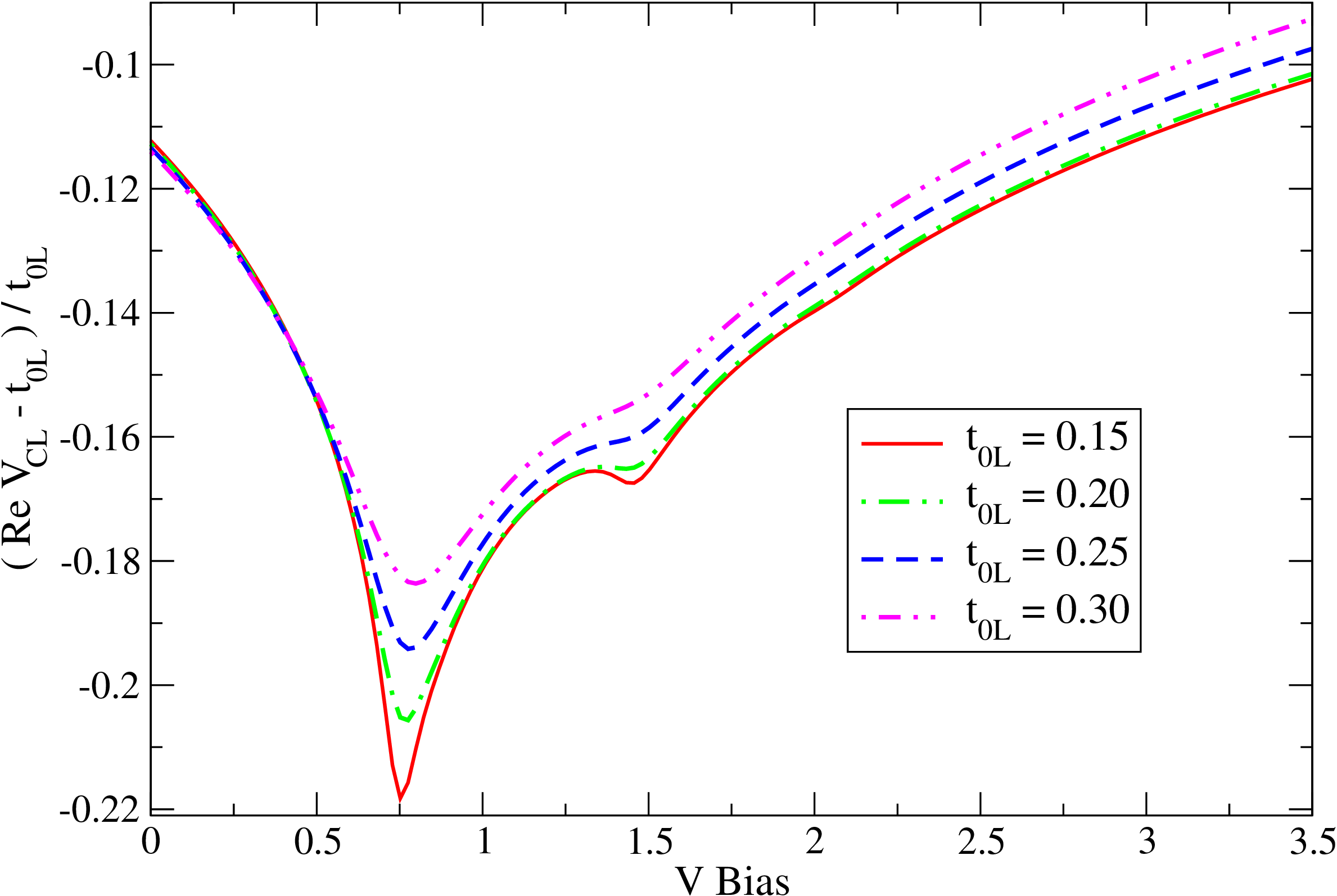}
  \caption{(Color online) Renormalised coupling $\tilde V_{CL}(V)$ for the off-resonant
  transport regime ($\varepsilon_0=0.50$) and $\gamma_A=0.14$ versus applied bias and 
  for different 
  values of the nominal hopping integrals $t_{0\alpha}$ ($\alpha=L,R$). The relative
  dependence of $\tilde V_{CL}$, shown here for $(\Re e \tilde V_{CL}(V) - t_{0\alpha})/t_{0\alpha}$,
  is linear only for small bias. Otherwise the dependence of $\tilde V_{CL}$ on both
  $t_{0\alpha}$ and $V$ is highly non-linear. The other parameters are the same as
  in Fig.~\ref{fig:eps050}.}
  \label{fig:VCL_varyt0L}
\end{figure}

We now turn onto the effect of the strength of the nominal hopping integral $t_{0L}$
on the renormalised coupling $\tilde V_{CL}(V)$ at fixed values of the crossing interaction 
strength $\gamma_A$. 
Figure \ref{fig:VCL_varyt0L} shows the relative dependence of $\tilde V_{CL}(V)$ on the nominal
coupling $t_{0L}$ at the left interface versus applied bias. The figure shows that there 
is no simple relationship between $\tilde V_{CL}(V)$ and $t_{0L}$ for the whole range of 
parameters explored. There is a small linear regime at low applied bias, otherwise the 
dependence of $\tilde V_{CL}$ on both $t_{0\alpha}$ and $V$ is highly non-linear.
There is a progressive washing-out of the features in $\tilde V_{CL}(V)$ for increasing 
values of $t_{0\alpha}$, since a general increase of the coupling to the leads 
generates to a global broadening of the features in the spectral functions.

\subsection{Non-equilibrium charge susceptibility}
\label{sec:NEChi}

In a recent paper \cite{Ness:2012}, we have developed the concept of the generalized
susceptibilities for nonlinear systems \cite{Safi:2011} and applied it to the charge
transport properties in two-terminal nano-devices. We have introduced the \NE 
charge susceptibility $\chi_c^{\rm NE}(V)$ concept, defined by 
$\chi_c^{\rm NE}(V)=\partial \langle n_C \rangle / \partial V$.
We have shown that $\chi_c^{\rm NE}(V)$ is related to the dynamical conductance $G(V)$. 
The relationship is formally different than the one obtained at equilibrium.
In spectroscopic terms, both $\chi_c^{\rm NE}(V)$ and $G(V)$ 
contain features versus applied bias when charge fluctuation occurs in the corresponding 
electronic resonance. This relationship has been demonstrated for model calculations
of interacting nanoscale devices but only when the interaction is present in the central
region \cite{Ness:2012}. 
We now check the validity of such a relationship between $\chi_c^{\rm NE}(V)$ 
and $G(V)$ when interaction crossing at the left contact is also taken into account.

\begin{figure}
  \includegraphics[width=72mm]{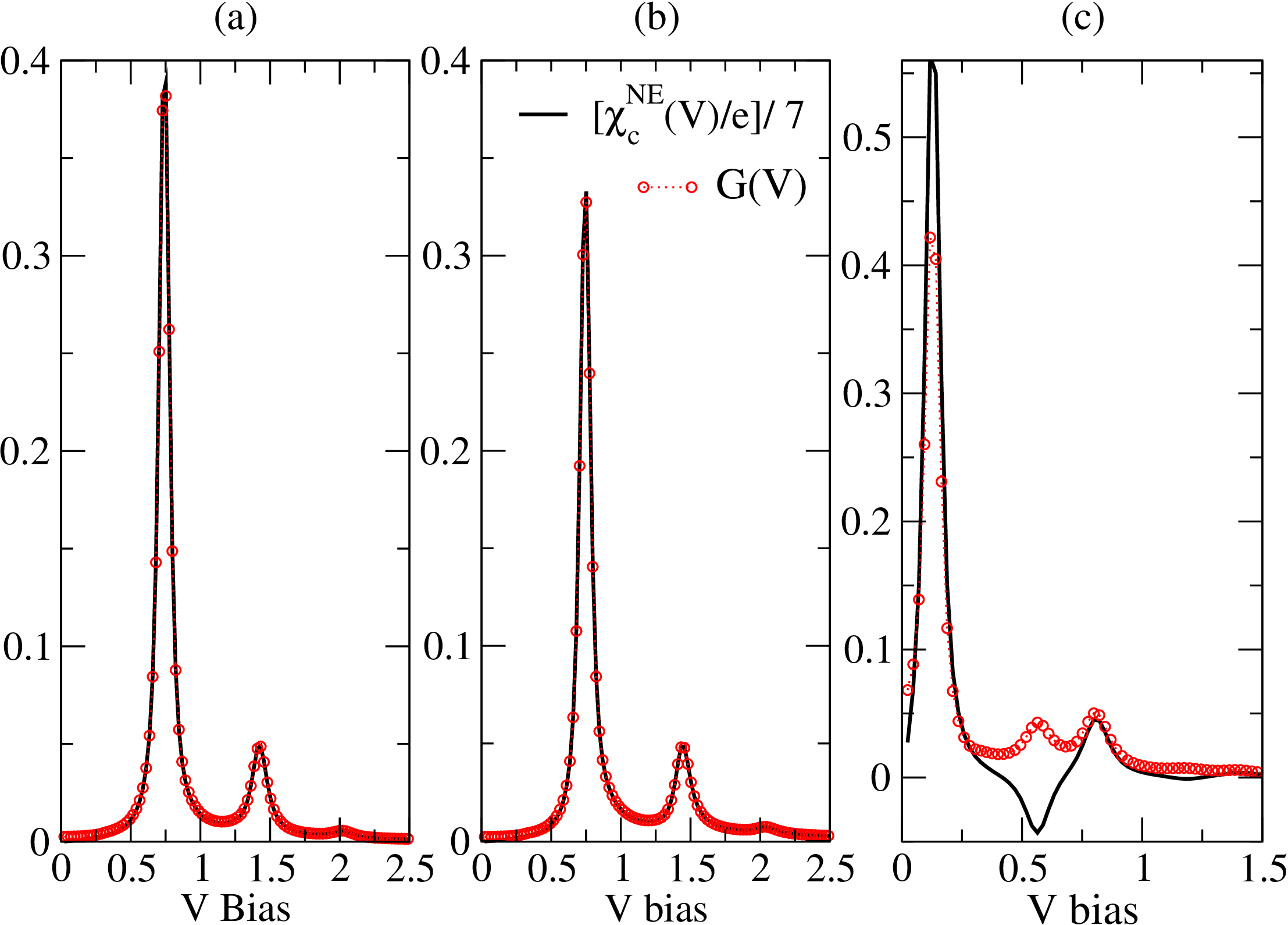}
  \caption{(Color online) Non-equilibrium charge susceptibility $\chi_c^{\rm NE}(V)$ and dynamical
  conductance $G(V)$ versus applied bias for different sets of parameters. $\chi_c^{\rm NE}(V)$
  is rescaled by a factor of 7.
  (a) Off-resonant regime $\varepsilon_0=0.5$ and $\gamma_A=0.08$.
  (b) Off-resonant regime $\varepsilon_0=0.5$ and $\gamma_A=0.14$,
  (c) Quasi-resonant regime $\varepsilon_0=0.2$ and $\gamma_A=0.05$.
  $\chi_c^{\rm NE}(V)$ and $G(V)$ present features at the same applied bias. Note
  the negative contribution to $\chi_c^{\rm NE}$ around $V \sim 0.56$. See main text 
  for a complete discussion.
  The other parameters are the same as in Fig.~\ref{fig:eps050}.}
  \label{fig:NEChiGV}
\end{figure}

Figure \ref{fig:NEChiGV} shows the results we obtain for $\chi_c^{\rm NE}(V)$ and $G(V)$ for
different sets of parameters corresponding to the off-resonant and quasi-resonant transport
regimes. Once more, we find that $\chi_c^{\rm NE}(V)$ and $G(V)$ present features at the same
applied bias. Such a behaviour validates the relationship between the non-equilibrium charge
susceptibility and the dynamical conductance even when interactions cross at the contacts (at
least at the static mean-field level).

Finally, one should note the negative contribution to $\chi_c^{\rm NE}(V)$ in panel (c) of
Figure \ref{fig:NEChiGV}. Such a behaviour originates from the properties of  $\chi_c^{\rm NE}$
and not from the approximation used to calculate the crossing interaction \cite{Ness:unpub2012}.
In fact, there are always two contributions to $\chi_c^{\rm NE}$, one is positive and corresponds
to electron fluctuation and the other is negative and corresponds to hole fluctuation. 
By electron (hole) fluctuations, we mean the variation of the occupancy (versus applied bias) of electronic resonances located nominally above (below) the Fermi level at equilibrium 
(or at small bias).
Therefore, for the off-resonant
transport regime with $\varepsilon_0 \gg \mu^{\rm eq}$, the features in $\chi_c^{\rm NE}(V)$ corresponds
to positive peaks (as shown in Panels (a) and (b) in Figure \ref{fig:NEChiGV} and in the different
figures of Ref.~[\onlinecite{Ness:2012}]). For the off-resonant regime with $\varepsilon_0 \ll 
\mu^{\rm eq}$, one would get negative peaks in $\chi_c^{\rm NE}(V)$. For the intermediate and 
quasi-resonant transport regime, one obtains both positive and negative contributions in 
$\chi_c^{\rm NE}(V)$ as shown in panel (c) of Figure \ref{fig:NEChiGV}.
The most peculiar case corresponds to a fully electron-hole symmetric system, i.e. the spectral
function $A_C(\omega)$ is fully symmetric around the equilibrium Fermi level and for any applied
bias $V$ with $\mu_{L,R}=\mu^{\rm eq} \pm eV/2$. In that case, one can easily show that
$\langle n_C \rangle = -i \int {\rm d}\omega\ G_C^<(\omega) / 2\pi$
is actually given by $\langle n_C \rangle \sim \int_{-\infty}^{+\infty} {\rm d}\omega\ A_C(\omega)$
for symmetry reasons. Hence $\langle n_C \rangle$ is independent of the applied bias for conserving
approximations for the self-energies, and consequently $\chi_c^{\rm NE}(V)=0$.
Such a behaviour can also be interpreted with the previous picture: for a fully electron-hole symmetric
system, any contributions from electron fluctuation is exactly cancelled out by the opposite 
contribution from hole fluctuation, and  $\chi_c^{\rm NE}(V)$ is flat and equal to zero 
for each bias \cite{Ness:unpub2012}.

\section{Conclusion}
\label{sec:ccl}

We have studied the transport properties through a two-terminal nanoscale device with interactions
present not only in the central region but also with interaction crossing at the interface between
the left lead and the central region. 
To calculate the current for such an fully interacting system, we have used our
recently developed quantum transport formula \cite{Ness:2011} based on the NEGF formalism.
As a first practical application, we have considered a prototypical single-molecule nanojunction
with electron-vibron interaction. In terms of the electron density matrix, the interaction is diagonal 
in the central region for the first vibron mode and off-diagonal between the central region
and the left electrode for the second vibron mode.
The interaction self-energies are calculated in a self-consistent manner using the lowest order
Hartree-Fock-like diagram in the central region and the Hartree-like diagram for the crossing 
interaction.
Our calculations were performed for different transport regimes ranging from the far off-resonance
to the quasi-resonant regime, and for a wide range of parameter values.
They shown that, for this model, we obtain a non-equilibrium (i.e. bias dependent)
static (i.e. energy independent) renormalisation of the nominal hopping matrix element between
the left electrode and the central region. The renormalisation is such that the amplitude of the
current is reduced in comparison with the current values obtained when the interaction is only 
present in the central region.
Such a result could provide an partial explanation for the fact in conventional density-functional 
based calculations, the values of the current are always much larger than in the corresponding 
experiments, since no non-equilibrium renormalisation of the contacts is taken into account
in those calculations.
However, even though it provides the right trends, the decrease in the current obtained by NE 
renormalisation of the coupling to the leads is not as important as
the effects obtained from a proper calculation of the band-gap and band-alignment in 
realistic molecular system \cite{Toher:2007,Thygesen:2008,Rostgaard:2010,Strange:2011}.

The NE static renormalisation of the contact is highly non-linear and non-monotonic in
function of the applied bias, and the larger effects occur at applied bias corresponding to
resonance peaks in the dynamical conductance. The conductance is also affected by the 
NE renormalisation of the contact, showing asymmetric broadening around the resonance 
peaks and some slight displacement of the peaks at large bias in function of the 
coupling strengh $\gamma_A$.

Furthermore, we have also shown that, even in the presence of crossing interactions, 
the relationship between the NE charge susceptibility and dynamical conductance \cite{Ness:2012} 
still holds for the different transport regimes considered here.

Extensions of the present study are now considered. One route is to develop
more accurate NE renormalisation by considering, for example, a quasi-particle approach 
within a dynamical mean-field-like treatement of the crossing interaction 
self-energy \cite{Ness:unpub2012}.
Another route is to go beyond the Hartree approximation for the crossing interaction 
self-energy using other many-body diagrams \cite{Dash:2011}. This would lead to 
dynamical NE renormalisation of the contact involving inelastic scattering processes. 

Finally, it should be noted that for the model system we considered here, one could also
solve the problem by using more standard approaches (typically the original Meir and 
Wingreen approach) by extending the size of the central region to include all the 
interaction. 
The results then obtained should be strictly equivalent to our calculations. 
We have already commented in detail on this point (in a formal and
theoretical point of view) in Refs.[\onlinecite{Ness:2012b,Ness:2011}].
However our approach offers a more intuitive physical
interpretation of the results, i.e. renormalisation of the contact
(in the present case in a static NE mean-field scheme). Using
an extended central region will not provide an easy physical 
interpretation of the results; and as a matter of principle,
it will not always be possible to increase at will the size 
of the central region, more especially when one considers future
application of the method to much large molecular system with
much more complex coupling to the leads.


\begin{thebibliography}{88}
\expandafter\ifx\csname natexlab\endcsname\relax\def\natexlab#1{#1}\fi
\expandafter\ifx\csname bibnamefont\endcsname\relax
  \def\bibnamefont#1{#1}\fi
\expandafter\ifx\csname bibfnamefont\endcsname\relax
  \def\bibfnamefont#1{#1}\fi
\expandafter\ifx\csname citenamefont\endcsname\relax
  \def\citenamefont#1{#1}\fi
\expandafter\ifx\csname url\endcsname\relax
  \def\url#1{\texttt{#1}}\fi
\expandafter\ifx\csname urlprefix\endcsname\relax\def\urlprefix{URL }\fi
\providecommand{\bibinfo}[2]{#2}
\providecommand{\eprint}[2][]{\url{#2}}

\bibitem[{\citenamefont{Widawsky et~al.}(2012)\citenamefont{Widawsky, Darancet,
  Neaton, and Venkataraman}}]{Widawsky:2012}
\bibinfo{author}{\bibfnamefont{J.~R.} \bibnamefont{Widawsky}},
  \bibinfo{author}{\bibfnamefont{P.}~\bibnamefont{Darancet}},
  \bibinfo{author}{\bibfnamefont{J.~B.} \bibnamefont{Neaton}},
  \bibnamefont{and}
  \bibinfo{author}{\bibfnamefont{L.}~\bibnamefont{Venkataraman}},
  \bibinfo{journal}{Nano Letters} \textbf{\bibinfo{volume}{12}},
  \bibinfo{pages}{354} (\bibinfo{year}{2012}).

\bibitem[{\citenamefont{Hirose and Tsukada}(1994)}]{Hirose:1994}
\bibinfo{author}{\bibfnamefont{K.}~\bibnamefont{Hirose}} \bibnamefont{and}
  \bibinfo{author}{\bibfnamefont{M.}~\bibnamefont{Tsukada}},
  \bibinfo{journal}{Physical Review Letters} \textbf{\bibinfo{volume}{73}},
  \bibinfo{pages}{150} (\bibinfo{year}{1994}).

\bibitem[{\citenamefont{DiVentra et~al.}(2000)\citenamefont{DiVentra,
  Pantelides, and Lang}}]{DiVentra:2000}
\bibinfo{author}{\bibfnamefont{M.}~\bibnamefont{DiVentra}},
  \bibinfo{author}{\bibfnamefont{S.~T.} \bibnamefont{Pantelides}},
  \bibnamefont{and} \bibinfo{author}{\bibfnamefont{N.~D.} \bibnamefont{Lang}},
  \bibinfo{journal}{Physical Review Letters} \textbf{\bibinfo{volume}{84}},
  \bibinfo{pages}{979} (\bibinfo{year}{2000}).

\bibitem[{\citenamefont{Taylor et~al.}(2001)\citenamefont{Taylor, Guo, and
  Wang}}]{Taylor:2001}
\bibinfo{author}{\bibfnamefont{J.}~\bibnamefont{Taylor}},
  \bibinfo{author}{\bibfnamefont{H.}~\bibnamefont{Guo}}, \bibnamefont{and}
  \bibinfo{author}{\bibfnamefont{J.}~\bibnamefont{Wang}},
  \bibinfo{journal}{Physical Review B} \textbf{\bibinfo{volume}{63}},
  \bibinfo{pages}{245407} (\bibinfo{year}{2001}).

\bibitem[{\citenamefont{Nardelli et~al.}(2001)\citenamefont{Nardelli,
  Fattebert, and Bernholc}}]{Nardelli:2001}
\bibinfo{author}{\bibfnamefont{M.~B.} \bibnamefont{Nardelli}},
  \bibinfo{author}{\bibfnamefont{J.-L.} \bibnamefont{Fattebert}},
  \bibnamefont{and} \bibinfo{author}{\bibfnamefont{J.}~\bibnamefont{Bernholc}},
  \bibinfo{journal}{Physical Review B} \textbf{\bibinfo{volume}{64}},
  \bibinfo{pages}{245423} (\bibinfo{year}{2001}).

\bibitem[{\citenamefont{Brandbyge et~al.}(2002)\citenamefont{Brandbyge, Mozos,
  Ordej\'on, Taylor, and Stokbro}}]{Brandbyge:2002}
\bibinfo{author}{\bibfnamefont{M.}~\bibnamefont{Brandbyge}},
  \bibinfo{author}{\bibfnamefont{J.-L.} \bibnamefont{Mozos}},
  \bibinfo{author}{\bibfnamefont{P.}~\bibnamefont{Ordej\'on}},
  \bibinfo{author}{\bibfnamefont{J.}~\bibnamefont{Taylor}}, \bibnamefont{and}
  \bibinfo{author}{\bibfnamefont{K.}~\bibnamefont{Stokbro}},
  \bibinfo{journal}{Physical Review B} \textbf{\bibinfo{volume}{65}},
  \bibinfo{pages}{165401} (\bibinfo{year}{2002}).

\bibitem[{\citenamefont{Gutierrez et~al.}(2002)\citenamefont{Gutierrez, Fagas,
  Cuniberti, Grossmann, Schmidt, and Richter}}]{Gutierrez:2002}
\bibinfo{author}{\bibfnamefont{R.}~\bibnamefont{Gutierrez}},
  \bibinfo{author}{\bibfnamefont{G.}~\bibnamefont{Fagas}},
  \bibinfo{author}{\bibfnamefont{G.}~\bibnamefont{Cuniberti}},
  \bibinfo{author}{\bibfnamefont{F.}~\bibnamefont{Grossmann}},
  \bibinfo{author}{\bibfnamefont{R.}~\bibnamefont{Schmidt}}, \bibnamefont{and}
  \bibinfo{author}{\bibfnamefont{K.}~\bibnamefont{Richter}},
  \bibinfo{journal}{Physical Review B} \textbf{\bibinfo{volume}{65}},
  \bibinfo{pages}{113410} (\bibinfo{year}{2002}).

\bibitem[{\citenamefont{Frauenheim et~al.}(2002)\citenamefont{Frauenheim,
  Seifert, Elstner, Niehaus, K\"ohler, Amkreutz, Sternberg, Hajnal, Carlo, and
  Suhai}}]{Frauenheim:2002}
\bibinfo{author}{\bibfnamefont{T.}~\bibnamefont{Frauenheim}},
  \bibinfo{author}{\bibfnamefont{G.}~\bibnamefont{Seifert}},
  \bibinfo{author}{\bibfnamefont{M.}~\bibnamefont{Elstner}},
  \bibinfo{author}{\bibfnamefont{T.}~\bibnamefont{Niehaus}},
  \bibinfo{author}{\bibfnamefont{C.}~\bibnamefont{K\"ohler}},
  \bibinfo{author}{\bibfnamefont{M.}~\bibnamefont{Amkreutz}},
  \bibinfo{author}{\bibfnamefont{M.}~\bibnamefont{Sternberg}},
  \bibinfo{author}{\bibfnamefont{Z.}~\bibnamefont{Hajnal}},
  \bibinfo{author}{\bibfnamefont{A.~D.} \bibnamefont{Carlo}}, \bibnamefont{and}
  \bibinfo{author}{\bibfnamefont{S.}~\bibnamefont{Suhai}},
  \bibinfo{journal}{Journal of Physics: Condensed Matter}
  \textbf{\bibinfo{volume}{14}}, \bibinfo{pages}{3015} (\bibinfo{year}{2002}).

\bibitem[{\citenamefont{Xue and Ratner}(2003)}]{Xue:2003}
\bibinfo{author}{\bibfnamefont{Y.}~\bibnamefont{Xue}} \bibnamefont{and}
  \bibinfo{author}{\bibfnamefont{M.~A.} \bibnamefont{Ratner}},
  \bibinfo{journal}{Physical Review B} \textbf{\bibinfo{volume}{68}},
  \bibinfo{pages}{115406} (\bibinfo{year}{2003}).

\bibitem[{\citenamefont{Louis et~al.}(2003)\citenamefont{Louis, Verg\'es,
  Palacios, P\'erez-Jim\'enez, and SanFabi\`an}}]{Louis+Palacios:2003}
\bibinfo{author}{\bibfnamefont{E.}~\bibnamefont{Louis}},
  \bibinfo{author}{\bibfnamefont{J.~A.} \bibnamefont{Verg\'es}},
  \bibinfo{author}{\bibfnamefont{J.~J.} \bibnamefont{Palacios}},
  \bibinfo{author}{\bibfnamefont{A.~J.} \bibnamefont{P\'erez-Jim\'enez}},
  \bibnamefont{and}
  \bibinfo{author}{\bibfnamefont{E.}~\bibnamefont{SanFabi\`an}},
  \bibinfo{journal}{Physical Review B} \textbf{\bibinfo{volume}{67}},
  \bibinfo{pages}{155321} (\bibinfo{year}{2003}).

\bibitem[{\citenamefont{Thygesen et~al.}(2003)\citenamefont{Thygesen,
  Bollinger, and Jacobsen}}]{Thygesen:2003}
\bibinfo{author}{\bibfnamefont{K.~S.} \bibnamefont{Thygesen}},
  \bibinfo{author}{\bibfnamefont{M.~V.} \bibnamefont{Bollinger}},
  \bibnamefont{and} \bibinfo{author}{\bibfnamefont{K.~W.}
  \bibnamefont{Jacobsen}}, \bibinfo{journal}{Physical Review B}
  \textbf{\bibinfo{volume}{67}}, \bibinfo{pages}{115404}
  (\bibinfo{year}{2003}).

\bibitem[{\citenamefont{Garc\'{\i}a-Su\'arez
  et~al.}(2005)\citenamefont{Garc\'{\i}a-Su\'arez, Rocha, Bailey, Lambert,
  Sanvito, and Ferrer}}]{Garcia-suarez:2005}
\bibinfo{author}{\bibfnamefont{V.~M.} \bibnamefont{Garc\'{\i}a-Su\'arez}},
  \bibinfo{author}{\bibfnamefont{A.~R.} \bibnamefont{Rocha}},
  \bibinfo{author}{\bibfnamefont{S.~W.} \bibnamefont{Bailey}},
  \bibinfo{author}{\bibfnamefont{C.~J.} \bibnamefont{Lambert}},
  \bibinfo{author}{\bibfnamefont{S.}~\bibnamefont{Sanvito}}, \bibnamefont{and}
  \bibinfo{author}{\bibfnamefont{J.}~\bibnamefont{Ferrer}},
  \bibinfo{journal}{Physical Review B} \textbf{\bibinfo{volume}{72}},
  \bibinfo{pages}{045437} (\bibinfo{year}{2005}).

\bibitem[{\citenamefont{Strange et~al.}(2011)\citenamefont{Strange, Rostgaard,
  H\"akkinen, and Thygesen}}]{Strange:2011}
\bibinfo{author}{\bibfnamefont{M.}~\bibnamefont{Strange}},
  \bibinfo{author}{\bibfnamefont{C.}~\bibnamefont{Rostgaard}},
  \bibinfo{author}{\bibfnamefont{H.}~\bibnamefont{H\"akkinen}},
  \bibnamefont{and} \bibinfo{author}{\bibfnamefont{K.~S.}
  \bibnamefont{Thygesen}}, \bibinfo{journal}{Physical Review B}
  \textbf{\bibinfo{volume}{83}}, \bibinfo{pages}{115108}
  (\bibinfo{year}{2011}).

\bibitem[{\citenamefont{Rangel et~al.}(2011)\citenamefont{Rangel, Ferretti,
  Trevisanutto, Olevano, and Rignanese}}]{Rangel:2011}
\bibinfo{author}{\bibfnamefont{T.}~\bibnamefont{Rangel}},
  \bibinfo{author}{\bibfnamefont{A.}~\bibnamefont{Ferretti}},
  \bibinfo{author}{\bibfnamefont{P.~E.} \bibnamefont{Trevisanutto}},
  \bibinfo{author}{\bibfnamefont{V.}~\bibnamefont{Olevano}}, \bibnamefont{and}
  \bibinfo{author}{\bibfnamefont{G.-M.} \bibnamefont{Rignanese}},
  \bibinfo{journal}{Phys. Rev. B} \textbf{\bibinfo{volume}{84}},
  \bibinfo{pages}{045426} (\bibinfo{year}{2011}).

\bibitem[{\citenamefont{Darancet et~al.}(2007)\citenamefont{Darancet, Ferretti,
  Mayou, and Olevano}}]{Darancet:2007}
\bibinfo{author}{\bibfnamefont{P.}~\bibnamefont{Darancet}},
  \bibinfo{author}{\bibfnamefont{A.}~\bibnamefont{Ferretti}},
  \bibinfo{author}{\bibfnamefont{D.}~\bibnamefont{Mayou}}, \bibnamefont{and}
  \bibinfo{author}{\bibfnamefont{V.}~\bibnamefont{Olevano}},
  \bibinfo{journal}{Phys. Rev. B} \textbf{\bibinfo{volume}{75}},
  \bibinfo{pages}{075102} (\bibinfo{year}{2007}).

\bibitem[{\citenamefont{Baym}(1962)}]{Baym:1962}
\bibinfo{author}{\bibfnamefont{G.}~\bibnamefont{Baym}},
  \bibinfo{journal}{Physical Review} \textbf{\bibinfo{volume}{127}},
  \bibinfo{pages}{1391} (\bibinfo{year}{1962}).

\bibitem[{\citenamefont{von Barth et~al.}(2005)\citenamefont{von Barth, Dahlen,
  van Leeuwen, and Stefanucci}}]{vonBarth:2005}
\bibinfo{author}{\bibfnamefont{U.}~\bibnamefont{von Barth}},
  \bibinfo{author}{\bibfnamefont{N.~E.} \bibnamefont{Dahlen}},
  \bibinfo{author}{\bibfnamefont{R.}~\bibnamefont{van Leeuwen}},
  \bibnamefont{and}
  \bibinfo{author}{\bibfnamefont{G.}~\bibnamefont{Stefanucci}},
  \bibinfo{journal}{Physical Review B} \textbf{\bibinfo{volume}{72}},
  \bibinfo{pages}{235109} (\bibinfo{year}{2005}).

\bibitem[{\citenamefont{van Leeuwen et~al.}(2006)\citenamefont{van Leeuwen,
  Dahlen, Stefanucci, Almbladh, and von Barth}}]{vanLeeuwen:2006}
\bibinfo{author}{\bibfnamefont{R.}~\bibnamefont{van Leeuwen}},
  \bibinfo{author}{\bibfnamefont{N.~E.} \bibnamefont{Dahlen}},
  \bibinfo{author}{\bibfnamefont{G.}~\bibnamefont{Stefanucci}},
  \bibinfo{author}{\bibfnamefont{C.-O.} \bibnamefont{Almbladh}},
  \bibnamefont{and} \bibinfo{author}{\bibfnamefont{U.}~\bibnamefont{von
  Barth}}, \bibinfo{journal}{Lecture Notes in Physics}
  \textbf{\bibinfo{volume}{706}}, \bibinfo{pages}{33} (\bibinfo{year}{2006}).

\bibitem[{\citenamefont{Kita}(2010)}]{Kita:2010}
\bibinfo{author}{\bibfnamefont{T.}~\bibnamefont{Kita}},
  \bibinfo{journal}{Progress of Theoretical Physics}
  \textbf{\bibinfo{volume}{123}}, \bibinfo{pages}{581} (\bibinfo{year}{2010}).

\bibitem[{\citenamefont{Tran}(2008)}]{Tran:2008}
\bibinfo{author}{\bibfnamefont{M.-T.} \bibnamefont{Tran}},
  \bibinfo{journal}{Physical Review B} \textbf{\bibinfo{volume}{78}},
  \bibinfo{pages}{125103} (\bibinfo{year}{2008}).

\bibitem[{\citenamefont{My\"oh\"anen et~al.}(2008)\citenamefont{My\"oh\"anen,
  Stan, Stefanucci, and van Leeuwen}}]{Myohanen:2008}
\bibinfo{author}{\bibfnamefont{P.}~\bibnamefont{My\"oh\"anen}},
  \bibinfo{author}{\bibfnamefont{A.}~\bibnamefont{Stan}},
  \bibinfo{author}{\bibfnamefont{G.}~\bibnamefont{Stefanucci}},
  \bibnamefont{and} \bibinfo{author}{\bibfnamefont{R.}~\bibnamefont{van
  Leeuwen}}, \bibinfo{journal}{EuroPhysics Letters}
  \textbf{\bibinfo{volume}{84}}, \bibinfo{pages}{67001} (\bibinfo{year}{2008}).

\bibitem[{\citenamefont{My\"oh\"anen et~al.}(2010)\citenamefont{My\"oh\"anen,
  Stan, Stefanucci, and van Leeuwen}}]{Myohanen:2010}
\bibinfo{author}{\bibfnamefont{P.}~\bibnamefont{My\"oh\"anen}},
  \bibinfo{author}{\bibfnamefont{A.}~\bibnamefont{Stan}},
  \bibinfo{author}{\bibfnamefont{G.}~\bibnamefont{Stefanucci}},
  \bibnamefont{and} \bibinfo{author}{\bibfnamefont{R.}~\bibnamefont{van
  Leeuwen}}, \bibinfo{journal}{Journal of Physics: Conference Series}
  \textbf{\bibinfo{volume}{220}}, \bibinfo{pages}{012017}
  (\bibinfo{year}{2010}).

\bibitem[{\citenamefont{Perfetto et~al.}(2010)\citenamefont{Perfetto,
  Stefanucci, and Cini}}]{Perfetto:2010}
\bibinfo{author}{\bibfnamefont{E.}~\bibnamefont{Perfetto}},
  \bibinfo{author}{\bibfnamefont{G.}~\bibnamefont{Stefanucci}},
  \bibnamefont{and} \bibinfo{author}{\bibfnamefont{M.}~\bibnamefont{Cini}},
  \bibinfo{journal}{Physical Review Letters} \textbf{\bibinfo{volume}{105}},
  \bibinfo{pages}{156802} (\bibinfo{year}{2010}).

\bibitem[{\citenamefont{Velick\'y et~al.}(2010)\citenamefont{Velick\'y,
  Kalvov\'a, and \v{S}pi\v{c}ka}}]{Velicky:2010}
\bibinfo{author}{\bibfnamefont{B.}~\bibnamefont{Velick\'y}},
  \bibinfo{author}{\bibfnamefont{A.}~\bibnamefont{Kalvov\'a}},
  \bibnamefont{and}
  \bibinfo{author}{\bibfnamefont{V.}~\bibnamefont{\v{S}pi\v{c}ka}},
  \bibinfo{journal}{Physical Review B} \textbf{\bibinfo{volume}{81}},
  \bibinfo{pages}{235116} (\bibinfo{year}{2010}).

\bibitem[{\citenamefont{von Friesen et~al.}(2009)\citenamefont{von Friesen,
  Verdozzi, and Almbladh}}]{PuigvonFriesen:2009}
\bibinfo{author}{\bibfnamefont{M.~P.} \bibnamefont{von Friesen}},
  \bibinfo{author}{\bibfnamefont{C.}~\bibnamefont{Verdozzi}}, \bibnamefont{and}
  \bibinfo{author}{\bibfnamefont{C.-O.} \bibnamefont{Almbladh}},
  \bibinfo{journal}{Physical Review Letters} \textbf{\bibinfo{volume}{103}},
  \bibinfo{pages}{176404} (\bibinfo{year}{2009}).

\bibitem[{\citenamefont{Arroyo et~al.}(2010)\citenamefont{Arroyo, Frederiksen,
  Rubio-Bollinger, V\'elez, Arnau, S\'anchez-Portal, and
  Agra\"{\i}t}}]{Arroyo:2010}
\bibinfo{author}{\bibfnamefont{C.~R.} \bibnamefont{Arroyo}},
  \bibinfo{author}{\bibfnamefont{T.}~\bibnamefont{Frederiksen}},
  \bibinfo{author}{\bibfnamefont{G.}~\bibnamefont{Rubio-Bollinger}},
  \bibinfo{author}{\bibfnamefont{M.}~\bibnamefont{V\'elez}},
  \bibinfo{author}{\bibfnamefont{A.}~\bibnamefont{Arnau}},
  \bibinfo{author}{\bibfnamefont{D.}~\bibnamefont{S\'anchez-Portal}},
  \bibnamefont{and}
  \bibinfo{author}{\bibfnamefont{N.}~\bibnamefont{Agra\"{\i}t}},
  \bibinfo{journal}{Physical Review B} \textbf{\bibinfo{volume}{81}},
  \bibinfo{pages}{075405} (\bibinfo{year}{2010}).

\bibitem[{\citenamefont{Ness and Fisher}(1999)}]{Ness:1999}
\bibinfo{author}{\bibfnamefont{H.}~\bibnamefont{Ness}} \bibnamefont{and}
  \bibinfo{author}{\bibfnamefont{A.~J.} \bibnamefont{Fisher}},
  \bibinfo{journal}{Physical Review Letters} \textbf{\bibinfo{volume}{83}},
  \bibinfo{pages}{452} (\bibinfo{year}{1999}).

\bibitem[{\citenamefont{Ness et~al.}(2001)\citenamefont{Ness, Shevlin, and
  Fisher}}]{Ness:2001}
\bibinfo{author}{\bibfnamefont{H.}~\bibnamefont{Ness}},
  \bibinfo{author}{\bibfnamefont{S.~A.} \bibnamefont{Shevlin}},
  \bibnamefont{and} \bibinfo{author}{\bibfnamefont{A.~J.}
  \bibnamefont{Fisher}}, \bibinfo{journal}{Physical Review B}
  \textbf{\bibinfo{volume}{63}}, \bibinfo{pages}{125422}
  (\bibinfo{year}{2001}).

\bibitem[{\citenamefont{Ness and Fisher}(2002)}]{Ness:2002a}
\bibinfo{author}{\bibfnamefont{H.}~\bibnamefont{Ness}} \bibnamefont{and}
  \bibinfo{author}{\bibfnamefont{A.~J.} \bibnamefont{Fisher}},
  \bibinfo{journal}{Europhysics Letters} \textbf{\bibinfo{volume}{57}},
  \bibinfo{pages}{885} (\bibinfo{year}{2002}).

\bibitem[{\citenamefont{Flensberg}(2003)}]{Flensberg:2003}
\bibinfo{author}{\bibfnamefont{K.}~\bibnamefont{Flensberg}},
  \bibinfo{journal}{Physical Review B} \textbf{\bibinfo{volume}{68}},
  \bibinfo{pages}{205323} (\bibinfo{year}{2003}).

\bibitem[{\citenamefont{Mii et~al.}(2003)\citenamefont{Mii, Tikhodeev, and
  Ueba}}]{Mii:2003}
\bibinfo{author}{\bibfnamefont{T.}~\bibnamefont{Mii}},
  \bibinfo{author}{\bibfnamefont{S.}~\bibnamefont{Tikhodeev}},
  \bibnamefont{and} \bibinfo{author}{\bibfnamefont{H.}~\bibnamefont{Ueba}},
  \bibinfo{journal}{Physical Review B} \textbf{\bibinfo{volume}{68}},
  \bibinfo{pages}{205406} (\bibinfo{year}{2003}).

\bibitem[{\citenamefont{Montgomery et~al.}(2003)\citenamefont{Montgomery,
  Hoekstra, Sutton, and Todorov}}]{Montgomery:2003b}
\bibinfo{author}{\bibfnamefont{M.~J.} \bibnamefont{Montgomery}},
  \bibinfo{author}{\bibfnamefont{J.}~\bibnamefont{Hoekstra}},
  \bibinfo{author}{\bibfnamefont{A.~P.} \bibnamefont{Sutton}},
  \bibnamefont{and} \bibinfo{author}{\bibfnamefont{T.~N.}
  \bibnamefont{Todorov}}, \bibinfo{journal}{Journal of Physics: Condensed
  Matter} \textbf{\bibinfo{volume}{15}}, \bibinfo{pages}{731}
  (\bibinfo{year}{2003}).

\bibitem[{\citenamefont{Troisi et~al.}(2003)\citenamefont{Troisi, Ratner, and
  Nitzan}}]{Troisi:2003}
\bibinfo{author}{\bibfnamefont{A.}~\bibnamefont{Troisi}},
  \bibinfo{author}{\bibfnamefont{M.~A.} \bibnamefont{Ratner}},
  \bibnamefont{and} \bibinfo{author}{\bibfnamefont{A.}~\bibnamefont{Nitzan}},
  \bibinfo{journal}{Journal of Chemical Physics}
  \textbf{\bibinfo{volume}{118}}, \bibinfo{pages}{6072} (\bibinfo{year}{2003}).

\bibitem[{\citenamefont{Chen et~al.}(2005{\natexlab{a}})\citenamefont{Chen,
  Zwolak, and di~Ventra}}]{Chen:2004}
\bibinfo{author}{\bibfnamefont{Y.~C.} \bibnamefont{Chen}},
  \bibinfo{author}{\bibfnamefont{M.}~\bibnamefont{Zwolak}}, \bibnamefont{and}
  \bibinfo{author}{\bibfnamefont{M.}~\bibnamefont{di~Ventra}},
  \bibinfo{journal}{Nano Letters} \textbf{\bibinfo{volume}{4}},
  \bibinfo{pages}{1709} (\bibinfo{year}{2005}{\natexlab{a}}).

\bibitem[{\citenamefont{Lorente and Persson}(2000)}]{Lorente:2000}
\bibinfo{author}{\bibfnamefont{N.}~\bibnamefont{Lorente}} \bibnamefont{and}
  \bibinfo{author}{\bibfnamefont{M.}~\bibnamefont{Persson}},
  \bibinfo{journal}{Physical Review Letters} \textbf{\bibinfo{volume}{85}},
  \bibinfo{pages}{2997} (\bibinfo{year}{2000}).

\bibitem[{\citenamefont{Frederiksen et~al.}(2004)\citenamefont{Frederiksen,
  Brandbyge, Lorente, and Jauho}}]{Frederiksen:2004}
\bibinfo{author}{\bibfnamefont{T.}~\bibnamefont{Frederiksen}},
  \bibinfo{author}{\bibfnamefont{M.}~\bibnamefont{Brandbyge}},
  \bibinfo{author}{\bibfnamefont{N.}~\bibnamefont{Lorente}}, \bibnamefont{and}
  \bibinfo{author}{\bibfnamefont{A.~P.} \bibnamefont{Jauho}},
  \bibinfo{journal}{Physical Review Letters} \textbf{\bibinfo{volume}{93}},
  \bibinfo{pages}{256601} (\bibinfo{year}{2004}).

\bibitem[{\citenamefont{Galperin
  et~al.}(2004{\natexlab{a}})\citenamefont{Galperin, Ratner, and
  Nitzan}}]{Galperin:2004}
\bibinfo{author}{\bibfnamefont{M.}~\bibnamefont{Galperin}},
  \bibinfo{author}{\bibfnamefont{M.~A.} \bibnamefont{Ratner}},
  \bibnamefont{and} \bibinfo{author}{\bibfnamefont{A.}~\bibnamefont{Nitzan}},
  \bibinfo{journal}{Nano Letters} \textbf{\bibinfo{volume}{4}},
  \bibinfo{pages}{1605} (\bibinfo{year}{2004}{\natexlab{a}}).

\bibitem[{\citenamefont{Galperin
  et~al.}(2004{\natexlab{b}})\citenamefont{Galperin, Ratner, and
  Nitzan}}]{Galperin:2004b}
\bibinfo{author}{\bibfnamefont{M.}~\bibnamefont{Galperin}},
  \bibinfo{author}{\bibfnamefont{M.~A.} \bibnamefont{Ratner}},
  \bibnamefont{and} \bibinfo{author}{\bibfnamefont{A.}~\bibnamefont{Nitzan}},
  \bibinfo{journal}{Journal of Chemical Physics}
  \textbf{\bibinfo{volume}{121}}, \bibinfo{pages}{11965}
  (\bibinfo{year}{2004}{\natexlab{b}}).

\bibitem[{\citenamefont{Mitra et~al.}(2004)\citenamefont{Mitra, Aleiner, and
  Millis}}]{Mitra:2004}
\bibinfo{author}{\bibfnamefont{A.}~\bibnamefont{Mitra}},
  \bibinfo{author}{\bibfnamefont{I.}~\bibnamefont{Aleiner}}, \bibnamefont{and}
  \bibinfo{author}{\bibfnamefont{A.~J.} \bibnamefont{Millis}},
  \bibinfo{journal}{Physical Review B} \textbf{\bibinfo{volume}{69}},
  \bibinfo{pages}{245302} (\bibinfo{year}{2004}).

\bibitem[{\citenamefont{Pecchia et~al.}(2004)\citenamefont{Pecchia, di~Carlo,
  Gagliardi, Sanna, Frauenhein, and Gutierrez}}]{Pecchia:2004}
\bibinfo{author}{\bibfnamefont{A.}~\bibnamefont{Pecchia}},
  \bibinfo{author}{\bibfnamefont{A.}~\bibnamefont{di~Carlo}},
  \bibinfo{author}{\bibfnamefont{A.}~\bibnamefont{Gagliardi}},
  \bibinfo{author}{\bibfnamefont{S.}~\bibnamefont{Sanna}},
  \bibinfo{author}{\bibfnamefont{T.}~\bibnamefont{Frauenhein}},
  \bibnamefont{and}
  \bibinfo{author}{\bibfnamefont{R.}~\bibnamefont{Gutierrez}},
  \bibinfo{journal}{Nano Letters} \textbf{\bibinfo{volume}{4}},
  \bibinfo{pages}{2109} (\bibinfo{year}{2004}).

\bibitem[{\citenamefont{Pecchia and di~Carlo}(2004)}]{Pecchia:2004b}
\bibinfo{author}{\bibfnamefont{A.}~\bibnamefont{Pecchia}} \bibnamefont{and}
  \bibinfo{author}{\bibfnamefont{A.}~\bibnamefont{di~Carlo}},
  \bibinfo{journal}{Reports on Progress in Physics}
  \textbf{\bibinfo{volume}{67}}, \bibinfo{pages}{1497} (\bibinfo{year}{2004}).

\bibitem[{\citenamefont{Chen et~al.}(2005{\natexlab{b}})\citenamefont{Chen,
  L\"u, and Zhu}}]{Chen_Z:2005}
\bibinfo{author}{\bibfnamefont{Z.}~\bibnamefont{Chen}},
  \bibinfo{author}{\bibfnamefont{R.}~\bibnamefont{L\"u}}, \bibnamefont{and}
  \bibinfo{author}{\bibfnamefont{B.}~\bibnamefont{Zhu}},
  \bibinfo{journal}{Physical Review B} \textbf{\bibinfo{volume}{71}},
  \bibinfo{pages}{165324} (\bibinfo{year}{2005}{\natexlab{b}}).

\bibitem[{\citenamefont{Paulsson et~al.}(2005)\citenamefont{Paulsson,
  Frederiksen, and Brandbyge}}]{Paulsson:2005}
\bibinfo{author}{\bibfnamefont{M.}~\bibnamefont{Paulsson}},
  \bibinfo{author}{\bibfnamefont{T.}~\bibnamefont{Frederiksen}},
  \bibnamefont{and}
  \bibinfo{author}{\bibfnamefont{M.}~\bibnamefont{Brandbyge}},
  \bibinfo{journal}{Physical Review B} \textbf{\bibinfo{volume}{72}},
  \bibinfo{pages}{201101} (\bibinfo{year}{2005}).

\bibitem[{\citenamefont{Ryndyk and Keller}(2005)}]{Ryndyk:2005}
\bibinfo{author}{\bibfnamefont{D.~A.} \bibnamefont{Ryndyk}} \bibnamefont{and}
  \bibinfo{author}{\bibfnamefont{J.}~\bibnamefont{Keller}},
  \bibinfo{journal}{Physical Review B} \textbf{\bibinfo{volume}{71}},
  \bibinfo{pages}{073305} (\bibinfo{year}{2005}).

\bibitem[{\citenamefont{Sergueev et~al.}(2005)\citenamefont{Sergueev, Roubtsov,
  and Guo}}]{Sergueev:2005}
\bibinfo{author}{\bibfnamefont{N.}~\bibnamefont{Sergueev}},
  \bibinfo{author}{\bibfnamefont{D.}~\bibnamefont{Roubtsov}}, \bibnamefont{and}
  \bibinfo{author}{\bibfnamefont{H.}~\bibnamefont{Guo}},
  \bibinfo{journal}{Physical Review Letters} \textbf{\bibinfo{volume}{95}},
  \bibinfo{pages}{146803} (\bibinfo{year}{2005}).

\bibitem[{\citenamefont{Viljas et~al.}(2005)\citenamefont{Viljas, Cuevas,
  Pauly, and H\"afner}}]{Viljas:2005}
\bibinfo{author}{\bibfnamefont{J.~K.} \bibnamefont{Viljas}},
  \bibinfo{author}{\bibfnamefont{J.~C.} \bibnamefont{Cuevas}},
  \bibinfo{author}{\bibfnamefont{F.}~\bibnamefont{Pauly}}, \bibnamefont{and}
  \bibinfo{author}{\bibfnamefont{M.}~\bibnamefont{H\"afner}},
  \bibinfo{journal}{Physical Review B} \textbf{\bibinfo{volume}{72}},
  \bibinfo{pages}{245415} (\bibinfo{year}{2005}).

\bibitem[{\citenamefont{Yamamoto et~al.}(2005)\citenamefont{Yamamoto, Watanabe,
  and Watanabe}}]{Yamamoto:2005}
\bibinfo{author}{\bibfnamefont{T.}~\bibnamefont{Yamamoto}},
  \bibinfo{author}{\bibfnamefont{K.}~\bibnamefont{Watanabe}}, \bibnamefont{and}
  \bibinfo{author}{\bibfnamefont{S.}~\bibnamefont{Watanabe}},
  \bibinfo{journal}{Physical Review Letters} \textbf{\bibinfo{volume}{95}},
  \bibinfo{pages}{065501} (\bibinfo{year}{2005}).

\bibitem[{\citenamefont{Cresti et~al.}(2006)\citenamefont{Cresti, Grosso, and
  Parravicini}}]{Cresti:2006}
\bibinfo{author}{\bibfnamefont{A.}~\bibnamefont{Cresti}},
  \bibinfo{author}{\bibfnamefont{G.}~\bibnamefont{Grosso}}, \bibnamefont{and}
  \bibinfo{author}{\bibfnamefont{G.~P.} \bibnamefont{Parravicini}},
  \bibinfo{journal}{Journal of Physics: Condensed Matter}
  \textbf{\bibinfo{volume}{18}}, \bibinfo{pages}{10059} (\bibinfo{year}{2006}).

\bibitem[{\citenamefont{Kula et~al.}(2006)\citenamefont{Kula, Jiang, and
  Luo}}]{Kula:2006}
\bibinfo{author}{\bibfnamefont{M.}~\bibnamefont{Kula}},
  \bibinfo{author}{\bibfnamefont{J.}~\bibnamefont{Jiang}}, \bibnamefont{and}
  \bibinfo{author}{\bibfnamefont{Y.}~\bibnamefont{Luo}}, \bibinfo{journal}{Nano
  Letters} \textbf{\bibinfo{volume}{6}}, \bibinfo{pages}{1693}
  (\bibinfo{year}{2006}).

\bibitem[{\citenamefont{Paulsson et~al.}(2006)\citenamefont{Paulsson,
  Frederiksen, and Brandbyge}}]{Paulsson:2006}
\bibinfo{author}{\bibfnamefont{M.}~\bibnamefont{Paulsson}},
  \bibinfo{author}{\bibfnamefont{T.}~\bibnamefont{Frederiksen}},
  \bibnamefont{and}
  \bibinfo{author}{\bibfnamefont{M.}~\bibnamefont{Brandbyge}},
  \bibinfo{journal}{Nano Letters} \textbf{\bibinfo{volume}{6}},
  \bibinfo{pages}{258} (\bibinfo{year}{2006}).

\bibitem[{\citenamefont{Ryndyk et~al.}(2006)\citenamefont{Ryndyk, Hartung, and
  Cuniberti}}]{Ryndyk:2006}
\bibinfo{author}{\bibfnamefont{D.~A.} \bibnamefont{Ryndyk}},
  \bibinfo{author}{\bibfnamefont{M.}~\bibnamefont{Hartung}}, \bibnamefont{and}
  \bibinfo{author}{\bibfnamefont{G.}~\bibnamefont{Cuniberti}},
  \bibinfo{journal}{Physical Review B} \textbf{\bibinfo{volume}{73}},
  \bibinfo{pages}{045420} (\bibinfo{year}{2006}).

\bibitem[{\citenamefont{Troisi and Ratner}(2006)}]{Troisi:2006b}
\bibinfo{author}{\bibfnamefont{A.}~\bibnamefont{Troisi}} \bibnamefont{and}
  \bibinfo{author}{\bibfnamefont{M.~A.} \bibnamefont{Ratner}},
  \bibinfo{journal}{Nano Letters} \textbf{\bibinfo{volume}{6}},
  \bibinfo{pages}{1784} (\bibinfo{year}{2006}).

\bibitem[{\citenamefont{de~la Vega et~al.}(2006)\citenamefont{de~la Vega,
  Mart\'{\i}n-Rodero, Agra\"{\i}t, and Levy-Yeyati}}]{Vega:2006}
\bibinfo{author}{\bibfnamefont{L.}~\bibnamefont{de~la Vega}},
  \bibinfo{author}{\bibfnamefont{A.}~\bibnamefont{Mart\'{\i}n-Rodero}},
  \bibinfo{author}{\bibfnamefont{N.}~\bibnamefont{Agra\"{\i}t}},
  \bibnamefont{and}
  \bibinfo{author}{\bibfnamefont{A.}~\bibnamefont{Levy-Yeyati}},
  \bibinfo{journal}{Physical Review B} \textbf{\bibinfo{volume}{73}},
  \bibinfo{pages}{075428} (\bibinfo{year}{2006}).

\bibitem[{\citenamefont{Toroker and Peskin}(2007)}]{Caspary:2007}
\bibinfo{author}{\bibfnamefont{M.~C.} \bibnamefont{Toroker}} \bibnamefont{and}
  \bibinfo{author}{\bibfnamefont{U.}~\bibnamefont{Peskin}},
  \bibinfo{journal}{Journal of Chemical Physics}
  \textbf{\bibinfo{volume}{127}}, \bibinfo{pages}{154706}
  (\bibinfo{year}{2007}).

\bibitem[{\citenamefont{Frederiksen et~al.}(2007)\citenamefont{Frederiksen,
  Paulsson, Brandbyge, and Jauho}}]{Frederiksen:2007}
\bibinfo{author}{\bibfnamefont{T.}~\bibnamefont{Frederiksen}},
  \bibinfo{author}{\bibfnamefont{M.}~\bibnamefont{Paulsson}},
  \bibinfo{author}{\bibfnamefont{M.}~\bibnamefont{Brandbyge}},
  \bibnamefont{and} \bibinfo{author}{\bibfnamefont{A.-P.} \bibnamefont{Jauho}},
  \bibinfo{journal}{Physical Review B} \textbf{\bibinfo{volume}{75}},
  \bibinfo{pages}{205413} (\bibinfo{year}{2007}).

\bibitem[{\citenamefont{Galperin et~al.}(2007)\citenamefont{Galperin, Nitzan,
  and Ratner}}]{Galperin:2007}
\bibinfo{author}{\bibfnamefont{M.}~\bibnamefont{Galperin}},
  \bibinfo{author}{\bibfnamefont{A.}~\bibnamefont{Nitzan}}, \bibnamefont{and}
  \bibinfo{author}{\bibfnamefont{M.~A.} \bibnamefont{Ratner}},
  \bibinfo{journal}{Physical Review B} \textbf{\bibinfo{volume}{74}},
  \bibinfo{pages}{075326} (\bibinfo{year}{2007}).

\bibitem[{\citenamefont{Ryndyk and Cuniberti}(2007)}]{Ryndyk:2007}
\bibinfo{author}{\bibfnamefont{D.~A.} \bibnamefont{Ryndyk}} \bibnamefont{and}
  \bibinfo{author}{\bibfnamefont{G.}~\bibnamefont{Cuniberti}},
  \bibinfo{journal}{Physical Review B} \textbf{\bibinfo{volume}{76}},
  \bibinfo{pages}{155430} (\bibinfo{year}{2007}).

\bibitem[{\citenamefont{Schmidt et~al.}(2007)\citenamefont{Schmidt, Hettler,
  and Sch\"on}}]{Schmidt:2007}
\bibinfo{author}{\bibfnamefont{B.~B.} \bibnamefont{Schmidt}},
  \bibinfo{author}{\bibfnamefont{M.~H.} \bibnamefont{Hettler}},
  \bibnamefont{and} \bibinfo{author}{\bibfnamefont{G.}~\bibnamefont{Sch\"on}},
  \bibinfo{journal}{Physical Review B} \textbf{\bibinfo{volume}{75}},
  \bibinfo{pages}{115125} (\bibinfo{year}{2007}).

\bibitem[{\citenamefont{Troisi et~al.}(2007)\citenamefont{Troisi, Beebe,
  Picraux, van Zee, Stewart, Ratner, and Kushmerick}}]{Troisi:2007}
\bibinfo{author}{\bibfnamefont{A.}~\bibnamefont{Troisi}},
  \bibinfo{author}{\bibfnamefont{J.~M.} \bibnamefont{Beebe}},
  \bibinfo{author}{\bibfnamefont{L.~B.} \bibnamefont{Picraux}},
  \bibinfo{author}{\bibfnamefont{R.~D.} \bibnamefont{van Zee}},
  \bibinfo{author}{\bibfnamefont{D.~R.} \bibnamefont{Stewart}},
  \bibinfo{author}{\bibfnamefont{M.~A.} \bibnamefont{Ratner}},
  \bibnamefont{and} \bibinfo{author}{\bibfnamefont{J.~G.}
  \bibnamefont{Kushmerick}}, \bibinfo{journal}{Proceedings of the National
  Academy of Sciences of the USA} \textbf{\bibinfo{volume}{104}},
  \bibinfo{pages}{14255} (\bibinfo{year}{2007}).

\bibitem[{\citenamefont{Asai}(2008)}]{Asai:2008}
\bibinfo{author}{\bibfnamefont{Y.}~\bibnamefont{Asai}},
  \bibinfo{journal}{Physical Review B} \textbf{\bibinfo{volume}{78}},
  \bibinfo{pages}{045434} (\bibinfo{year}{2008}).

\bibitem[{\citenamefont{Benesch et~al.}(2008)\citenamefont{Benesch,
  \v{C}\'{\i}\v{z}ek, Klime\v{s}, Kondov, Thoss, and Domcke}}]{Benesch:2008}
\bibinfo{author}{\bibfnamefont{C.}~\bibnamefont{Benesch}},
  \bibinfo{author}{\bibfnamefont{M.}~\bibnamefont{\v{C}\'{\i}\v{z}ek}},
  \bibinfo{author}{\bibfnamefont{J.}~\bibnamefont{Klime\v{s}}},
  \bibinfo{author}{\bibfnamefont{I.}~\bibnamefont{Kondov}},
  \bibinfo{author}{\bibfnamefont{M.}~\bibnamefont{Thoss}}, \bibnamefont{and}
  \bibinfo{author}{\bibfnamefont{W.}~\bibnamefont{Domcke}},
  \bibinfo{journal}{Journal of Physical Chemistry C}
  \textbf{\bibinfo{volume}{112}}, \bibinfo{pages}{9880} (\bibinfo{year}{2008}).

\bibitem[{\citenamefont{Paulsson et~al.}(2008)\citenamefont{Paulsson,
  Frederiksen, Ueba, Lorente, and Brandbyge}}]{Paulsson:2008}
\bibinfo{author}{\bibfnamefont{M.}~\bibnamefont{Paulsson}},
  \bibinfo{author}{\bibfnamefont{T.}~\bibnamefont{Frederiksen}},
  \bibinfo{author}{\bibfnamefont{H.}~\bibnamefont{Ueba}},
  \bibinfo{author}{\bibfnamefont{N.}~\bibnamefont{Lorente}}, \bibnamefont{and}
  \bibinfo{author}{\bibfnamefont{M.}~\bibnamefont{Brandbyge}},
  \bibinfo{journal}{Physical Review Letters} \textbf{\bibinfo{volume}{100}},
  \bibinfo{pages}{226604} (\bibinfo{year}{2008}).

\bibitem[{\citenamefont{Egger and Gogolin}(2008)}]{Egger:2008}
\bibinfo{author}{\bibfnamefont{R.}~\bibnamefont{Egger}} \bibnamefont{and}
  \bibinfo{author}{\bibfnamefont{A.~O.} \bibnamefont{Gogolin}},
  \bibinfo{journal}{Physical Review B} \textbf{\bibinfo{volume}{77}},
  \bibinfo{pages}{113405} (\bibinfo{year}{2008}).

\bibitem[{\citenamefont{Monturet and Lorente}(2008)}]{Monturet:2008}
\bibinfo{author}{\bibfnamefont{S.}~\bibnamefont{Monturet}} \bibnamefont{and}
  \bibinfo{author}{\bibfnamefont{N.}~\bibnamefont{Lorente}},
  \bibinfo{journal}{Physical Review B} \textbf{\bibinfo{volume}{78}},
  \bibinfo{pages}{035445} (\bibinfo{year}{2008}).

\bibitem[{\citenamefont{McEniry et~al.}(2008)\citenamefont{McEniry,
  Frederiksen, Todorov, Dundas, and Horsfield}}]{McEniry:2008}
\bibinfo{author}{\bibfnamefont{E.~J.} \bibnamefont{McEniry}},
  \bibinfo{author}{\bibfnamefont{T.}~\bibnamefont{Frederiksen}},
  \bibinfo{author}{\bibfnamefont{T.~N.} \bibnamefont{Todorov}},
  \bibinfo{author}{\bibfnamefont{D.}~\bibnamefont{Dundas}}, \bibnamefont{and}
  \bibinfo{author}{\bibfnamefont{A.~P.} \bibnamefont{Horsfield}},
  \bibinfo{journal}{Physical Review B} \textbf{\bibinfo{volume}{78}},
  \bibinfo{pages}{035446} (\bibinfo{year}{2008}).

\bibitem[{\citenamefont{Ryndyk et~al.}(2009)\citenamefont{Ryndyk, Gutiérrez,
  Song, and Cuniberti}}]{Ryndyk:2008}
\bibinfo{author}{\bibfnamefont{D.~A.} \bibnamefont{Ryndyk}},
  \bibinfo{author}{\bibfnamefont{R.}~\bibnamefont{Gutiérrez}},
  \bibinfo{author}{\bibfnamefont{B.}~\bibnamefont{Song}}, \bibnamefont{and}
  \bibinfo{author}{\bibfnamefont{G.}~\bibnamefont{Cuniberti}}, in
  \emph{\bibinfo{booktitle}{Energy Transfer Dynamics in Biomaterial Systems}},
  edited by \bibinfo{editor}{\bibfnamefont{I.}~\bibnamefont{Burghardt}},
  \bibinfo{editor}{\bibfnamefont{V.}~\bibnamefont{May}},
  \bibinfo{editor}{\bibfnamefont{D.~A.} \bibnamefont{Micha}}, \bibnamefont{and}
  \bibinfo{editor}{\bibfnamefont{E.~R.} \bibnamefont{Bittner}}
  (\bibinfo{publisher}{Springer Berlin Heidelberg}, \bibinfo{year}{2009}),
  vol.~\bibinfo{volume}{93} of \emph{\bibinfo{series}{Springer Series in
  Chemical Physics}}, pp. \bibinfo{pages}{213--335}, ISBN
  \bibinfo{isbn}{978-3-642-02306-4}.

\bibitem[{\citenamefont{Schmidt et~al.}(2008)\citenamefont{Schmidt, Hettler,
  and Sch\"{o}n}}]{Schmidt:2008}
\bibinfo{author}{\bibfnamefont{B.~B.} \bibnamefont{Schmidt}},
  \bibinfo{author}{\bibfnamefont{M.~H.} \bibnamefont{Hettler}},
  \bibnamefont{and}
  \bibinfo{author}{\bibfnamefont{G.}~\bibnamefont{Sch\"{o}n}},
  \bibinfo{journal}{Physical Review B (Condensed Matter and Materials Physics)}
  \textbf{\bibinfo{volume}{77}}, \bibinfo{pages}{165337}
  (\bibinfo{year}{2008}).

\bibitem[{\citenamefont{Tsukada and Mitsutake}(2009)}]{Tsukada:2009}
\bibinfo{author}{\bibfnamefont{M.}~\bibnamefont{Tsukada}} \bibnamefont{and}
  \bibinfo{author}{\bibfnamefont{K.}~\bibnamefont{Mitsutake}},
  \bibinfo{journal}{Journal of the Physical Society of Japan}
  \textbf{\bibinfo{volume}{78}}, \bibinfo{pages}{084701}
  (\bibinfo{year}{2009}).

\bibitem[{\citenamefont{Loos et~al.}(2009)\citenamefont{Loos, Koch, Alvermann,
  Bishop, and Fehske}}]{Loos:2009}
\bibinfo{author}{\bibfnamefont{J.}~\bibnamefont{Loos}},
  \bibinfo{author}{\bibfnamefont{T.}~\bibnamefont{Koch}},
  \bibinfo{author}{\bibfnamefont{A.}~\bibnamefont{Alvermann}},
  \bibinfo{author}{\bibfnamefont{A.~R.} \bibnamefont{Bishop}},
  \bibnamefont{and} \bibinfo{author}{\bibfnamefont{H.}~\bibnamefont{Fehske}},
  \bibinfo{journal}{Journal of Physics: Condensed Matter}
  \textbf{\bibinfo{volume}{21}}, \bibinfo{pages}{395601}
  (\bibinfo{year}{2009}).

\bibitem[{\citenamefont{Secker et~al.}(2011)\citenamefont{Secker, Wagner,
  Ballmann, Hartle, Thoss, and Weber}}]{Secker:2011}
\bibinfo{author}{\bibfnamefont{D.}~\bibnamefont{Secker}},
  \bibinfo{author}{\bibfnamefont{S.}~\bibnamefont{Wagner}},
  \bibinfo{author}{\bibfnamefont{S.}~\bibnamefont{Ballmann}},
  \bibinfo{author}{\bibfnamefont{R.}~\bibnamefont{Hartle}},
  \bibinfo{author}{\bibfnamefont{M.}~\bibnamefont{Thoss}}, \bibnamefont{and}
  \bibinfo{author}{\bibfnamefont{H.~B.} \bibnamefont{Weber}},
  \bibinfo{journal}{Physical Review Letters} \textbf{\bibinfo{volume}{106}},
  \bibinfo{pages}{136807} (\bibinfo{year}{2011}).

\bibitem[{\citenamefont{H\"artle and Thoss}(2011)}]{Hartle:2011a}
\bibinfo{author}{\bibfnamefont{R.}~\bibnamefont{H\"artle}} \bibnamefont{and}
  \bibinfo{author}{\bibfnamefont{M.}~\bibnamefont{Thoss}},
  \bibinfo{journal}{Physical Review B} \textbf{\bibinfo{volume}{83}},
  \bibinfo{pages}{115414} (\bibinfo{year}{2011}).

\bibitem[{\citenamefont{H\"artle et~al.}(2011)\citenamefont{H\"artle, Butzin,
  Rubio-Pons, and Thoss}}]{Hartle:2011b}
\bibinfo{author}{\bibfnamefont{R.}~\bibnamefont{H\"artle}},
  \bibinfo{author}{\bibfnamefont{M.}~\bibnamefont{Butzin}},
  \bibinfo{author}{\bibfnamefont{O.}~\bibnamefont{Rubio-Pons}},
  \bibnamefont{and} \bibinfo{author}{\bibfnamefont{M.}~\bibnamefont{Thoss}},
  \bibinfo{journal}{Physical Review Letters} \textbf{\bibinfo{volume}{107}},
  \bibinfo{pages}{046802} (\bibinfo{year}{2011}).

\bibitem[{\citenamefont{Ness and Dash}(2011)}]{Ness:2011}
\bibinfo{author}{\bibfnamefont{H.}~\bibnamefont{Ness}} \bibnamefont{and}
  \bibinfo{author}{\bibfnamefont{L.}~\bibnamefont{Dash}},
  \bibinfo{journal}{Physical Review B} \textbf{\bibinfo{volume}{84}},
  \bibinfo{pages}{235428} (\bibinfo{year}{2011}).

\bibitem[{\citenamefont{Ness and Dash}(2012{\natexlab{a}})}]{Ness:2012b}
\bibinfo{author}{\bibfnamefont{H.}~\bibnamefont{Ness}} \bibnamefont{and}
  \bibinfo{author}{\bibfnamefont{L.}~\bibnamefont{Dash}},
  \bibinfo{journal}{Journal of Physics A: Mathematical and Theoretical}
  \textbf{\bibinfo{volume}{45}}, \bibinfo{pages}{195301}
  (\bibinfo{year}{2012}{\natexlab{a}}).

\bibitem[{\citenamefont{Meir and Wingreen}(1992)}]{Meir:1992}
\bibinfo{author}{\bibfnamefont{Y.}~\bibnamefont{Meir}} \bibnamefont{and}
  \bibinfo{author}{\bibfnamefont{N.~S.} \bibnamefont{Wingreen}},
  \bibinfo{journal}{Physical Review Letters} \textbf{\bibinfo{volume}{68}},
  \bibinfo{pages}{2512} (\bibinfo{year}{1992}).

\bibitem[{\citenamefont{Perfetto et~al.}(2012)\citenamefont{Perfetto,
  Stefanucci, and Cini}}]{Perfetto:2012}
\bibinfo{author}{\bibfnamefont{E.}~\bibnamefont{Perfetto}},
  \bibinfo{author}{\bibfnamefont{G.}~\bibnamefont{Stefanucci}},
  \bibnamefont{and} \bibinfo{author}{\bibfnamefont{M.}~\bibnamefont{Cini}},
  \bibinfo{journal}{Physical Review B} \textbf{\bibinfo{volume}{85}},
  \bibinfo{pages}{165437} (\bibinfo{year}{2012}).

\bibitem[{\citenamefont{Ness and Dash}(2012{\natexlab{b}})}]{Ness:2012}
\bibinfo{author}{\bibfnamefont{H.}~\bibnamefont{Ness}} \bibnamefont{and}
  \bibinfo{author}{\bibfnamefont{L.}~\bibnamefont{Dash}},
  \bibinfo{journal}{Physical Review Letters} \textbf{\bibinfo{volume}{108}},
  \bibinfo{pages}{126401} (\bibinfo{year}{2012}{\natexlab{b}}).

\bibitem[{\citenamefont{Dash et~al.}(2010)\citenamefont{Dash, Ness, and
  Godby}}]{Dash:2010}
\bibinfo{author}{\bibfnamefont{L.~K.} \bibnamefont{Dash}},
  \bibinfo{author}{\bibfnamefont{H.}~\bibnamefont{Ness}}, \bibnamefont{and}
  \bibinfo{author}{\bibfnamefont{R.~W.} \bibnamefont{Godby}},
  \bibinfo{journal}{Journal of Chemical Physics}
  \textbf{\bibinfo{volume}{132}}, \bibinfo{pages}{104113}
  (\bibinfo{year}{2010}).

\bibitem[{\citenamefont{Ness et~al.}(2010)\citenamefont{Ness, Dash, and
  Godby}}]{Ness:2010}
\bibinfo{author}{\bibfnamefont{H.}~\bibnamefont{Ness}},
  \bibinfo{author}{\bibfnamefont{L.}~\bibnamefont{Dash}}, \bibnamefont{and}
  \bibinfo{author}{\bibfnamefont{R.~W.} \bibnamefont{Godby}},
  \bibinfo{journal}{Physical Review B} \textbf{\bibinfo{volume}{82}},
  \bibinfo{pages}{085426} (\bibinfo{year}{2010}).

\bibitem[{\citenamefont{Dash et~al.}(2011)\citenamefont{Dash, Ness, and
  Godby}}]{Dash:2011}
\bibinfo{author}{\bibfnamefont{L.~K.} \bibnamefont{Dash}},
  \bibinfo{author}{\bibfnamefont{H.}~\bibnamefont{Ness}}, \bibnamefont{and}
  \bibinfo{author}{\bibfnamefont{R.~W.} \bibnamefont{Godby}},
  \bibinfo{journal}{Physical Review B} \textbf{\bibinfo{volume}{84}},
  \bibinfo{pages}{085433} (\bibinfo{year}{2011}).

\bibitem[{\citenamefont{Heeger et~al.}(1988)\citenamefont{Heeger, Kivelson,
  Schrieffer, and Su}}]{Heeger:1988}
\bibinfo{author}{\bibfnamefont{A.~J.} \bibnamefont{Heeger}},
  \bibinfo{author}{\bibfnamefont{S.}~\bibnamefont{Kivelson}},
  \bibinfo{author}{\bibfnamefont{J.~R.} \bibnamefont{Schrieffer}},
  \bibnamefont{and} \bibinfo{author}{\bibfnamefont{W.-P.} \bibnamefont{Su}},
  \bibinfo{journal}{Review of Modern Physics} \textbf{\bibinfo{volume}{60}},
  \bibinfo{pages}{781} (\bibinfo{year}{1988}).

\bibitem[{\citenamefont{Datta et~al.}(1997)\citenamefont{Datta, Tian, Hong,
  Reifenberger, Henderson, and Kubiak}}]{Datta:1997}
\bibinfo{author}{\bibfnamefont{S.}~\bibnamefont{Datta}},
  \bibinfo{author}{\bibfnamefont{W.~D.} \bibnamefont{Tian}},
  \bibinfo{author}{\bibfnamefont{S.~H.} \bibnamefont{Hong}},
  \bibinfo{author}{\bibfnamefont{R.}~\bibnamefont{Reifenberger}},
  \bibinfo{author}{\bibfnamefont{J.~I.} \bibnamefont{Henderson}},
  \bibnamefont{and} \bibinfo{author}{\bibfnamefont{C.~P.}
  \bibnamefont{Kubiak}}, \bibinfo{journal}{Physical Review Letters}
  \textbf{\bibinfo{volume}{79}}, \bibinfo{pages}{2530} (\bibinfo{year}{1997}).

\bibitem[{\citenamefont{L.~K.~Dash and Godby}(2012)}]{Dash:2012}
\bibinfo{author}{\bibfnamefont{M.~V.} \bibnamefont{L.~K.~Dash},
  \bibfnamefont{H.~Ness}} \bibnamefont{and}
  \bibinfo{author}{\bibfnamefont{R.~W.} \bibnamefont{Godby}},
  \bibinfo{journal}{Journal of Chemical Physics}
  \textbf{\bibinfo{volume}{136}}, \bibinfo{pages}{064708}
  (\bibinfo{year}{2012}).

\bibitem[{\citenamefont{Safi and Joyez}(2011)}]{Safi:2011}
\bibinfo{author}{\bibfnamefont{I.}~\bibnamefont{Safi}} \bibnamefont{and}
  \bibinfo{author}{\bibfnamefont{P.}~\bibnamefont{Joyez}},
  \bibinfo{journal}{Physical Review B} \textbf{\bibinfo{volume}{84}},
  \bibinfo{pages}{205129} (\bibinfo{year}{2011}).

\bibitem{Ness:unpub2012}
H. Ness and L. K. Dash, unpublished.

\bibitem[{\citenamefont{Toher and Sanvito}(2007)}]{Toher:2007}
\bibinfo{author}{\bibfnamefont{C.}~\bibnamefont{Toher}} \bibnamefont{and}
  \bibinfo{author}{\bibfnamefont{S.}~\bibnamefont{Sanvito}},
  \bibinfo{journal}{Physical Review Letters} \textbf{\bibinfo{volume}{99}},
  \bibinfo{pages}{056801} (\bibinfo{year}{2007}).

\bibitem[{\citenamefont{Thygesen}(2008)}]{Thygesen:2008}
\bibinfo{author}{\bibfnamefont{K.~S.} \bibnamefont{Thygesen}},
  \bibinfo{journal}{Physical Review Letters} \textbf{\bibinfo{volume}{100}},
  \bibinfo{pages}{166804} (\bibinfo{year}{2008}).

\bibitem[{\citenamefont{Rostgaard et~al.}(2010)\citenamefont{Rostgaard,
  Jacobsen, and Thygesen}}]{Rostgaard:2010}
\bibinfo{author}{\bibfnamefont{C.}~\bibnamefont{Rostgaard}},
  \bibinfo{author}{\bibfnamefont{K.~W.} \bibnamefont{Jacobsen}},
  \bibnamefont{and} \bibinfo{author}{\bibfnamefont{K.~S.}
  \bibnamefont{Thygesen}}, \bibinfo{journal}{Physical Review B}
  \textbf{\bibinfo{volume}{81}}, \bibinfo{pages}{085103}
  (\bibinfo{year}{2010}).

\end{thebibliography}
\end{document}